\documentclass[twocolumn]{aastex631}

\usepackage[caption=false]{subfig}
\usepackage{graphicx}	% Including figure files
\usepackage{amsmath}	% Advanced maths commands
\usepackage{booktabs}
\usepackage{multirow}

\begin{document}

%\title{The Variable Asymmetry of PSR J1622-0315 Over Five Years}
\title{The Orbit and Companion of PSR J1622-0315: Variable Asymmetry and a Massive Neutron Star}
\shorttitle{The Orbit and Companion of PSR J1622-0315}

\correspondingauthor{Bidisha Sen}
\email{bidisha.sen@ntnu.no}

\author[0000-0002-1845-9325]{Bidisha Sen}
\affiliation{Department of Physics, Norwegian University of Science and Technology \\
NO-7491 Trondheim, Norway}

\author[0000-0002-0237-1636]{Manuel Linares}
\affiliation{Department of Physics, Norwegian University of Science and Technology \\
NO-7491 Trondheim, Norway}
\affiliation{Departament de Física, EEBE, Universitat Politècnica de Catalunya \\
Av. Eduard Maristany 16, E-08019 Barcelona, Spain}

\author[0000-0001-6894-6044]{Mark R. Kennedy}
\affiliation{Jodrell Bank Centre for Astrophysics, Department of Physics and Astronomy \\
The University of Manchester, Manchester M13 9PL,
United Kingdom}
\affiliation{School of Physics, University College Cork \\
Cork, Ireland}

\author[0000-0001-8522-4983]{Rene P. Breton}
\affiliation{Jodrell Bank Centre for Astrophysics, Department of Physics and Astronomy \\
The University of Manchester, Manchester M13 9PL,
United Kingdom}

\author[0000-0003-4260-960X]{Devina Misra}
\affiliation{Department of Physics, Norwegian University of Science and Technology \\
NO-7491 Trondheim, Norway}

\author[0000-0003-0438-4956]{Marco Turchetta}
\affiliation{Department of Physics, Norwegian University of Science and Technology \\
NO-7491 Trondheim, Norway}

\author[0000-0003-4236-9642]{Vikram S. Dhillon}
\affiliation{Department of Physics and Astronomy, University of Sheffield \\
Sheffield S3 7RH, UK}
\affiliation{Instituto de Astrofísica de Canarias \\
E-38205 La Laguna, Tenerife, Spain}

\author[0000-0003-0245-9424]{Daniel Mata S\'anchez}
\affiliation{Instituto de Astrofísica de Canarias \\
E-38205 La Laguna, Tenerife, Spain}
\affiliation{Departamento de Astrofísica, Universidad de La Laguna \\
E-38206 La Laguna, Tenerife, Spain}

\author[0000-0003-4355-3572]{Colin J. Clark}
\affiliation{Max Planck Institute for Gravitational Physics (Albert Einstein Institute) \\
D-30167 Hannover, Germany}
\affiliation{Leibniz Universität Hannover \\
D-30167 Hannover, Germany}

%% Note that the \and command from previous versions of AASTeX is now
%% depreciated in this version as it is no longer necessary. AASTeX 
%% automatically takes care of all commas and "and"s between authors names.

%% AASTeX 6.31 has the new \collaboration and \nocollaboration commands to
%% provide the collaboration status of a group of authors. These commands 
%% can be used either before or after the list of corresponding authors. The
%% argument for \collaboration is the collaboration identifier. Authors are
%% encouraged to surround collaboration identifiers with ()s. The 
%% \nocollaboration command takes no argument and exists to indicate that
%% the nearby authors are not part of surrounding collaborations.

%% Mark off the abstract in the ``abstract'' environment. 
\begin{abstract}
The companion to PSR J1622-0315, one of the most compact known redback millisecond pulsars, shows extremely low irradiation despite its short orbital period. We model this system to determine the binary parameters, combining optical observations from NTT in 2017 and NOT in 2022 with the binary modeling code \textsc{ICARUS}. We find a best-fit neutron star mass of $2.3 \pm 0.4\,\text{M}_\odot $, and a companion mass of $0.15 \pm 0.02\,\text{M}_\odot$. We detect for the first time low-level irradiation from asymmetry in the minima as well as a change in the asymmetry of the maxima of its light curves over five years. Using star spot models, we find better fits than those from symmetric direct heating models, with consistent orbital parameters. We discuss an alternative scenario where the changing asymmetry is produced by a variable intrabinary shock. In summary, we find that PSR J1622-0315 combines low irradiation with variable light curve asymmetry, and a relatively high neutron star mass.
% 250 words limit, currently 158

\end{abstract}

%% Keywords should appear after the \end{abstract} command. 
%% The AAS Journals now uses Unified Astronomy Thesaurus concepts:
%% https://astrothesaurus.org
%% You will be asked to selected these concepts during the submission process
%% but this old "keyword" functionality is maintained in case authors want
%% to include these concepts in their preprints.

%% From the front matter, we move on to the body of the paper.
%% Sections are demarcated by \section and \subsection, respectively.
%% Observe the use of the LaTeX \label
%% command after the \subsection to give a symbolic KEY to the
%% subsection for cross-referencing in a \ref command.
%% You can use LaTeX's \ref and \label commands to keep track of
%% cross-references to sections, equations, tables, and figures.
%% That way, if you change the order of any elements, LaTeX will
%% automatically renumber them.
%%
%% We recommend that authors also use the natbib \citep
%% and \citet commands to identify citations.  The citations are
%% tied to the reference list via symbolic KEYs. The KEY corresponds
%% to the KEY in the \bibitem in the reference list below. 

\section{Introduction}
Compact binary millisecond pulsar systems are promising sources to find the most massive neutron stars.
%, which have accreted mass from their gravitationally bound low-mass companion star \citep{massivensbw}. This duration of recycling spins up the pulsars to millisecond spin periods, with masses most likely greater than that of pulsars with slower spin periods 
Accretion of mass from their gravitationally bound companions spin up the pulsars to millisecond spin periods \citep[recycling scenario;][]{recycling}. As a result, millisecond pulsars are expected to be more massive than those with slower spin periods \citep{spinup1, spinup2}. Among the binaries hosting millisecond pulsars, those nicknamed as ``spiders" have revealed particularly high NS masses (eg. \citealp{Linares18b}).
%(eg. \citealp{massivensbw}, but see also \citealp{Clark2023}).
%These ``spider" systems, nicknamed as such because the relativistic pulsar wind ablates the companion 
Spiders are characterised by low-mass companion stars ablated by a relativistic pulsar wind, due to their small orbital separations. They are classified as ``redbacks" (RBs) if the companion has a mass of order $0.1\,\text{M}_{\sun}$, or ``black widows" (BWs) if the companion has a mass of order $0.01\,\text{M}_{\sun}$ \citep{Fruchter1988AMP, Damico01, Roberts2012}. %When the companion is close to filling its Roche lobe, the star distorts into a tear drop shape, with the nose of the star facing the pulsar. 
The high energy particles in the pulsar wind often heat the nose of the distorted, Roche-lobe filling companion, irradiating the day side of the star \citep{Kluzniak1998}. 
%While irradiation can be a dominant feature in the light curves of these systems, some spider systems have light curves whose dominant features are caused by the geometry of the companion star and the orbit. In particular, 
Heavily irradiated systems show optical light curves with only one maximum per orbit, at superior conjunction of the companion star when the day side is maximally visible. However, when there is little irradiation and the companion is filling (or close to filling) its Roche-lobe, ellipsoidal modulation from the distorted, tear-drop shaped surface of the companion star dominates. The optical light curve of such a system shows two maxima per orbit, located at the quadratures of the orbit, when the companion is seen sideways. 

One system exhibiting the latter features in its light curves is PSR J1622-0315, henceforth referred to as J1622. This RB system was discovered via radio observations with the Robert C. Byrd Green Bank Telescope while searching for unassociated \textit{Fermi}-LAT sources. It has a short orbital period of 3.9 hours with a close to overflowing Roche lobe \citep{Sanpa16}, making it an extremely compact system \citep[see Table 5 of][]{Strader19}, and is thought to have one of the lightest known RB companions \citep{Strader19, Y23}. The X-ray spectrum taken with \textit{XMM-Newton} shows hard X-ray emission that fits a power law of $\Gamma = 2.0 \pm 0.3$ well, but there are not enough counts to constrain its orbital variability \citep{Gentile2018}. Analysis of $\gamma$-ray observations with \textit{Fermi}-LAT did not detect any eclipses, constraining the inclination of the system to below $83.4^\circ$ \citep{Clark2023}, and revealed gamma ray pulsations \citep{3pc}. 

Optical spectroscopic analysis by \cite{Strader19} revealed a minimum pulsar mass of $1.45\pm0.08$\,M$_\odot$, with a companion mass of $0.10-0.14$\,M$_\odot$, along with radial velocity curves that imply a radial velociy semi-amplitude for the companion of $423 \pm 8\,\text{km s}^{-1}$. Optical light curves were taken by \cite{Y23} (hereby referred to as Y23) with the Lulin $1$\,m telescope and Lijiang $2.4$\,m telescope, which led to a calculated pulsar mass of $1.84 \pm 0.19\,\text{M}_\odot$, and a companion of mass $0.122_{-0.006}^{+0.007}\,\text{M}_\odot$, under the assumption of no irradiation \citep{Y23}. The optical light curves from \cite{Turchetta23} show flat colors over the orbit in the \textit{g'}, \textit{r'}, and \textit{i'} bands, further supporting the claim that irradiation is not significant in this system.

% While the multiband light curves indicate that ellipsoidal modulation is the dominating effect and the color diagrams indicate little irradiation \citep{Turchetta23}, we observe that there is a small amount of variable irradiation in this system, which can be seen over a five year time span. 

Here we present new rapid multi-band photometry of J1622 and the results of detailed modelling of optical light curves taken in 2017 and 2022. We observe light curves with asymmetric minima and maxima, that we model with symmetric, direct heating models that take gravity darkening and irradiation into account. We also use asymmetric models that include flux differences from adding star spots on top of the direct heating model. In Section \ref{datasets} we explain the observations analyzed in this work. In Section \ref{modeling} we present our modeling code and methods. We present our results in Section \ref{results} and discuss them in Section \ref{discussion}. 

\section{Observations and Data Analysis} \label{datasets}
\subsection{NTT/ULTRACAM: The 2017 campaign}
J1622 was observed using ULTRACAM \citep{Dhillon07} mounted in the European Southern Observatory's 3.5 m New Technology Telescope (NTT) over three consecutive nights in June of 2017. ULTRACAM provides three optical filters of data simultaneously, with a readout time of 24 ms. The data from 2017 June 17 and 18 were taken using the $u_{\rm s}$, $g_{\rm s}$, and $i_{\rm s}$ Super Sloan Digital Sky Survey (Super-SDSS) filters, while the data from 2017 June 19 were taken using $u_{\rm s}$, $g_{\rm s}$, and $r_{\rm s}$. These filters cover the same wavelength ranges as the traditional SDSS filters \citep{Doi2010}, but with a higher throughput \citep{Dhillon21}. On-chip binning of 2x2 was used for data on 2017 June 17 
%due to clouds during the observations
to improve the signal-to-noise ratio (SNR) and compensate for the presence of clouds during the observations, and 1x1 binning for data taken on 2017 June 18-19. The exposure times for each individual frame in the $g_{\rm s}$, $r_{\rm s}$, and $i_{\rm s}$ filters were 13 s. ULTRACAM allows for readout cycles to be skipped 
%for $u_{\rm s}$ frames in order to increase the SNR of the data
for $u_{\rm s}$ frames in order to increase the effective exposure time, and therefore, the SNR. For the data from 2017 June 17, 5 cycles were skipped, leading to an effective exposure time of 65 s for each $u_{\rm s}$ frame, while 3 cycles were skipped for 2017 June 18-19, leading to effective exposure times of 39 s for these data.

The data were bias-corrected and flat-fielded using calibration frames taken on each night of the observations, and aperture photometry performed for J1622 and 3 nearby, stable comparison stars \footnote{PanSTARRS IDs: 104062457272005214, 104102457539119045, 104052457386904888}. All of these tasks were performed using the HiPERCAM pipeline \citep{Dhillon21}. Aperture photometry was performed for a series of aperture sizes, ranging from $0.4\times \textrm{PSF}_{\rm FWHM}$ up to $2.0\times \textrm{PSF}_{\rm FWHM}$. The counts for the target and each reference star then were extracted using the aperture size which maximised the SNR. For the target star, this was typically between 0.7 and 0.8 the measured FWHM, while for the reference stars, this was 1.0-1.1 $\times$ the FWHM. These fluxes were corrected to an aperture of infinite radius using the Curve of Growth method \citep{Howell90}. Night-to-night variations in the zero-point and changes in the transmission of the atmosphere were corrected using ``ensemble'' photometry \citep{Honeycutt92}.

\begin{figure*}
    \centering
    \includegraphics[trim={6.2cm 0cm 8cm 0cm}, clip, scale=0.26]{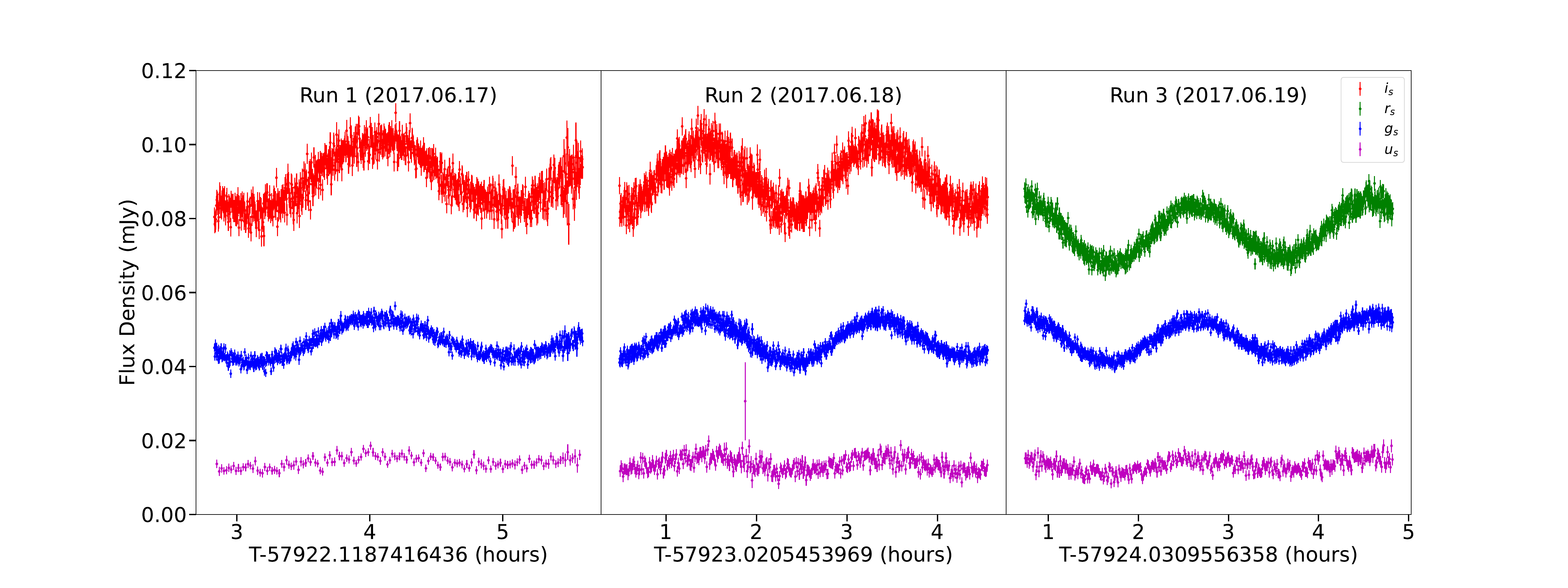}
    \caption{ Simultaneous ULTRACAM light curves of J1622 for our three nights of observations in the $u_s, g_s, r_s, i_s$\, filters. The first night is shown on the leftmost panel, the middle night is on the middle panel, and the last night is on the rightmost panel. The second and third nights have full phase coverage of the system. Light curves are available as supplementary online data. }
    \label{fig:ucam_data}
\end{figure*}

The magnitudes of the target in each filter were calibrated against the SDSS magnitudes of the nearby reference stars. The resulting J1622 light curves are shown in Fig. \ref{fig:ucam_data}. We binned this data by orbital phase into 50 bins and used the mean magnitude of each bin to model the light curves for computational efficiency with our MCMC code. The uncertainty for each data point in a bin were propagated to give the uncertainty of the bin.

\subsection{NOT/ALFOSC Data: 2022 Observations}
We observed J1622 using the Alhambra Faint Object Spectrograph and Camera (ALFOSC) camera mounted on the $2.56$\,m Nordic Optical Telescope (NOT) during the night of 2022/04/21. The optical images were acquired in a $6.4\,\arcmin\times6.4\,\arcmin$ field-of-view centered on the target, with $2$\,min long exposures alternated between the SDSS \textit{g'}, \textit{r'} and \textit{i'} filters for four consecutive hours. We used $2\times2$ binning for the CCD to reduce the readout times down to $8.1$\,s per exposure.

%We processed our data with bias subtraction and flat field correction using \textsc{IRAF}\footnote{\url{https://iraf-community.github.io/}} routines. 

Optical light curves derived from these observations were presented in \cite{Turchetta23}, with identical reduction procedures. We employed the \textsc{ULTRACAM} \citep{Dhillon07}
% \footnote{\url{https://cygnus.astro.warwick.ac.uk/phsaap/software/ultracam/html/index.html}} 
software package to perform differential aperture photometry of J1622, setting the aperture radius to 2 times the seeing and using the same set of three stable comparison stars selected for the NTT/ULTRACAM photometry. This set of reference stars showed very low variability in the NOT data as well (rms amplitudes $\simeq0.010\text{, } 0.005 \text{, and } 0.006\text{\,mag}$ for the \textit{g'}, \textit{r'}, and \textit{i'} filters respectively). 

\subsection{SOAR Optical Spectra}\label{soar}

\begin{figure}
    \centering
    %trim={0cm 0cm 0cm 0cm}, clip,
    \includegraphics[scale=0.35]{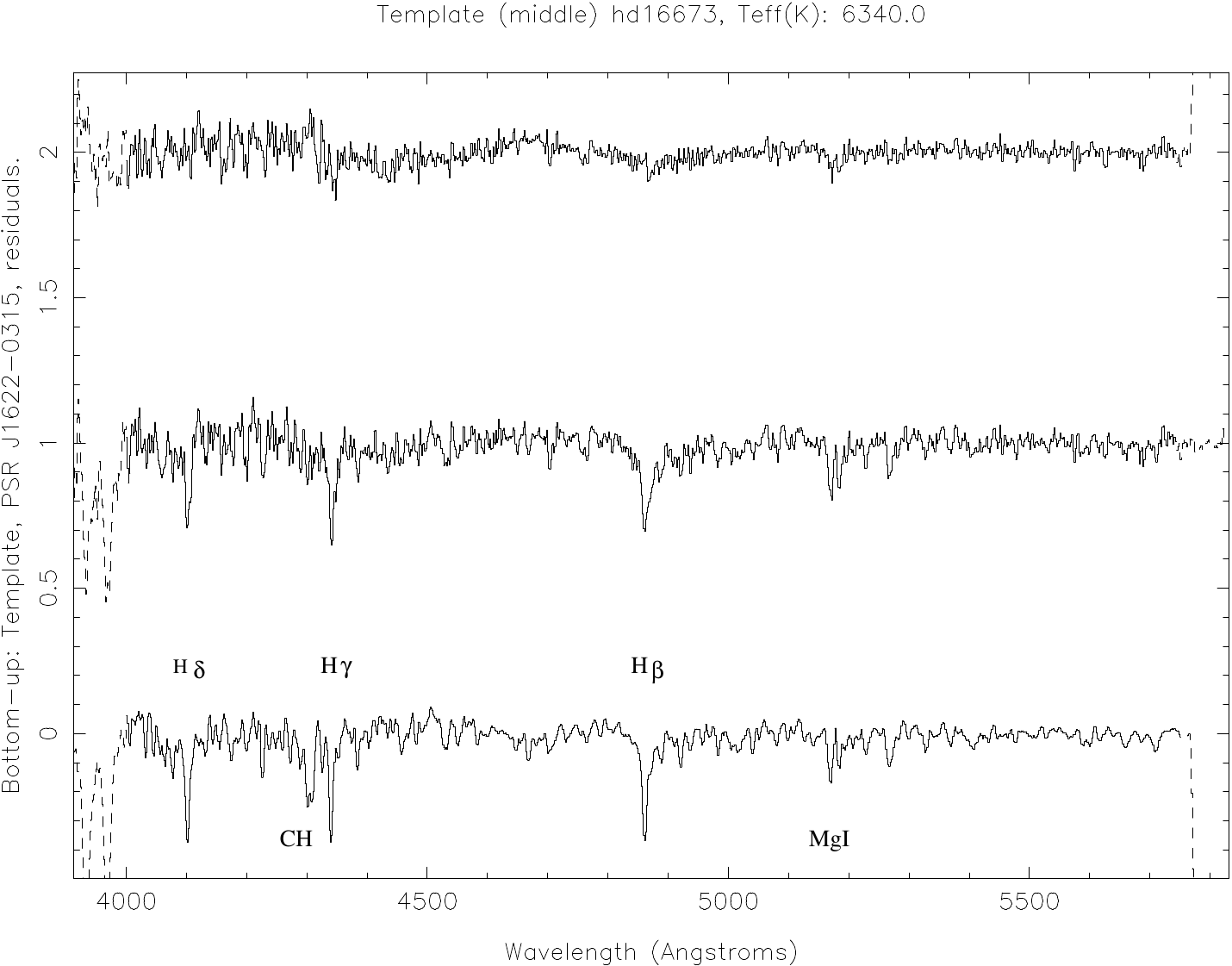}
    \caption{{\it From bottom to top:} best-fit F6 spectral template (T$_\text{eff}$=6340~K), average spectrum of J1622 and residuals from optimal subtraction.}
    \label{fig:optsub_spectra}
\end{figure}

We re-analyzed the Southern Astrophysical Research Telescope (SOAR) optical spectra of J1622 presented by
\citep[][see reduction and extraction procedures in their Section 2]{Strader19}, in order to derive independent constraints on the
temperature of the companion star.
We applied the so-called optimal subtraction method to compare the
relative strength of absorption lines in these spectra with a set of
standard stars with known effective temperatures, $T_\text{eff}$ \citep{Marsh94}.
To that end, we used a set of templates from
UVES-POP \citep{Bagnulo03} (see Appendix A in \citealt{Linares18b} for
details) within the spectral analysis package
\textsc{molly}, degraded to the spectral resolution of the J1622 SOAR
spectra (250 km s$^{-1}$).
We shifted the normalized spectra in velocity to the reference frame
of the companion star, using the ephemerides and radial velocity curve
fit from \cite[][$K_2 = 423 \text{km s}^{-1}$ and systemic velocity
$\gamma = -135 \text{km s}^{-1}$]{Strader19}.
We then averaged the 10 spectra with the highest signal-to-noise
ratio, which covered all orbital phases except 0.4-0.6 (i.e., except
superior conjunction of the companion star).
We also repeated this analysis using only 2 spectra taken around phase
0, and found consistent results (same $T_\text{eff}$ within the errors, see Section \ref{optsub}).

\section{\textsc{ICARUS} and Binary Modelling}\label{modeling}
% For the NOT data, we use the conversion outlined in the SDSS DR7 Photometric Flux Calibration documentation \footnote{\url{https://classic.sdss.org/dr7/algorithms/fluxcal.php\#sdss2flux}}.
To model the light curves, we used the stellar binary light curve synthesis code \textsc{ICARUS} \citep{Breton2012}. We first converted the apparent AB magnitudes of each dataset into flux densities using the corresponding zero-point flux densities: 3631 Jy for the SDSS and Super SDSS filters \textit{u'}, \textit{g'}, \textit{r'}, and \textit{i'}. These flux densities were provided as an input into \textsc{ICARUS}, alongside theoretical specific intensities for each filter which span a range of effective temperatures, log surface gravities, and viewing angle. The grids of specific intensities were created by integrating the product of atmosphere grids generated using \textsc{ATLAS9} \citep{Kurucz2004} and the transmission curve for each filter (Super SDSS for NTT and SDSS for NOT). We define the beginning of the orbit $\phi_\text{orb} = 0$ at inferior conjunction of the companion.
% While the NOT dataset has phase data with $\phi=0$ corresponding to the inferior conjunction of the companion, the NTT dataset follows the radio phase convention of $\phi=0.25$ corresponding to inferior conjunction. To account for this, we add a phase shift of $0.25$ to the NTT dataset when supplying ICARUS with the data.

\subsection{\textsc{ICARUS} Parameters}\label{icarus}
\textsc{ICARUS} models the tessellated surface of the companion star in a binary system by solving the Roche potential equation and combining this with pre-computed stellar atmosphere grids. It then calculates the emitted flux to create model light curves at all orbital phases as would be observed at a particular distance and viewing angle from the system. We have three static parameters that we leave fixed, with eight free direct heating parameters (accounting for gravity darkening and irradiation) and four star spot parameters that we vary and sample using Markov Chain Monte Carlo (MCMC) algorithms, as explained in Section \ref{MCMC}. We assume a tidally locked system and fix the co-rotation parameter, $\omega$, to $1$. We also set the gravity darkening coefficient, $\beta$, to $0.08$. This value corresponds to a convective envelope for the companion \citep{Lucy1967} and is in line with a low mass companion star.
% , which we assume given that the companion is lighter than $3M\sun$ \cite{3M}. Both datasets show a lack of asymmetry in the minima that would indicate a radiative envelope for the companion, which would require higher choice of $\beta = 0.25$ \cite{VZ1,VZ2}.
The last fixed parameter is the orbital period $P_\text{orb}$, which we set to the \cite{Sanpa16} reported value of $0.1617006798\pm6\times10^{-10}$ days.

We derive one of the required input parameters, the mass ratio $q$, from the semi-major axis of the pulsar $x_1$ and the projected radial velocity semi-amplitude of the companion $K_2$. First, we use the relation:

\begin{equation}
    K_1 = \frac{2 \pi x}{P_\text{orb}}
\end{equation}
in order to find the projected radial velocity semi-amplitude of the pulsar, $K_1$. We use this in the following relation:

\begin{equation}
    q = \frac{K_2}{K_1} = \frac{M_1}{M_2}\text{,}
\end{equation}
where the index $1$ indicates the pulsar and $2$ is the companion, to find $q$. The filling factor used in \textsc{ICARUS} is the ratio of the distance from the center mass of the companion to its nose $r_\text{nose}$, to the distance from the center of mass to the $L_1$ Lagrange point $r_{L_1}$:

\begin{equation}\label{fillr}
    f = \frac{r_\text{nose}}{r_{L_1}}\text{.}
\end{equation}
The other five direct heating parameters include the inclination angle $i$, which is the angle between the orbital angular momentum of the system and the line of sight, the base temperature before applying gravity darkening $T_\text{base}$, irradiation temperature $T_\text{irr}$, distance modulus m-M, and the extinction in the Johnson V band $A_v$.

When adding star spots, \textsc{ICARUS} models each spot as a 2D Gaussian, where the location of the spot on the surface of the star is given by two angles $\theta_\text{spot}$ and $\phi_\text{spot}$ (polar and azimuthal, respectively). The spread of the spot is controlled by the spot radius  $R_\text{spot}$, which is the standard deviation of the 2D Gaussian, given in radians. The last parameter of the spot is its temperature difference $T_\text{spot}$, which is positive for hot spots and negative for cold spots. The flux from a star spot is added to or subtracted from the flux after accounting for gravity darkening and irradiation.

% We note that the relation between $\phi_\text{orbit}$ and $\phi_\text{spot}$ is given by:

% \begin{equation*}
%     \phi_\text{orbit} = \frac{\phi_\text{spot} + \pi}{2 \pi} \text{.}
% \end{equation*}

Finally, \textsc{ICARUS} requires a band calibration uncertainty that captures any model atmosphere uncertainties by allowing for 
%error
independent offsets on each modeled band. For both datasets, we set the band uncertainty to 0.01 mag. When a model light curve is compared with an inputted dataset, \textsc{ICARUS} calculates a best-fit magnitude offset, to better match the model light curve to the data. This offset is done with a linear fitting algorithm and accounts for variations in the absolute calibration of the photometry. 

\subsection{Prior Distributions}\label{priors}
We apply uniform priors to $x_1$, $f$, $T_\text{irr}$, and m-M. We use the \cite{Sanpa16} reported value of $0.219258\pm5\times10^{-6}$\,lt-s as the bounds of the flat distribution for $x_1$, which was derived from their measurement of the semimajor axis of the pulsar, $a_1$. The filling factor uses a flat prior between $0$ and $1$, while we place an upper bound on the irradiation temperature at $5000$\,K given the constant colors observed in \cite{Turchetta23}. For m-M, we use the GAIA DR1, 2, and 3 parallax estimates of J1622 \citep{gaiadr1a, gaiadr1b, gaiadr1archive, gaiadr2, gaiadr3, gaiadr3catval} to find the bounds of the distribution. Since edge-on systems are more likely to be detected, we model $\cos{i}$ with a uniform distribution, corresponding to an isotropic distribution of $i$ after accounting for projection effects on the sky plane. Since no gamma ray eclipse was detected in J1622, we place an upper limit on $i$ of $83.4^{\circ}$ \citep{Clark2023}. For the star spot models, we apply uniform priors on the temperature and radius of the spot.

The rest of the parameters are assigned Gaussian priors. We use the mean and three times the standard deviation values reported in \cite{Strader19} from optical spectroscopy to have a conservative range on the prior for $K_2$, since the reported $1\sigma$ values only take statistical uncertainties into account. Using the same spectroscopic data, we perform optimal subtraction to provide constraints on the effective temperature $T_\text{eff}$ in Sections \ref{soar} and \ref{optsub}. From there, we derive a Gaussian prior on $T_\text{base}$ of $6400 \pm 250$\,K. For $A_V$, we use the \cite{Green2019} dust maps in order to get a color excess E(g$-$r) $=0.23 \pm 0.02$, which was estimated using a reddening law of $R_V = 3.32 \pm 0.18$ \citep{rv}. Following the discussion outlined in the Bayestar19 usage notes, we first convert the E(g$-$r) from the dust maps to E(g$-$r) in Pan-STARRS 1 passbands using the relation from \cite{Bayestar17}:

\begin{equation}
    E(g-r)_{P1} = 0.901 \cdot E(g-r)_\text{Bayestar19}\text{.} \\
\end{equation}

Next, we convert this to E(B$-$V) using the relation from \cite{Sf11}:
\begin{equation}
    E(B-V) = 0.981 \pm 0.02 \cdot E(g-r) \text{.}
\end{equation}
Putting these two relations together, we find that E(B$-$V) for our source is $0.203 \pm 0.018$. We convert this to get an estimate of the extinction: $A_V = 0.67 \pm 0.07$. We also apply a Gaussian prior on $\theta_\text{spot}$ around the equator since star spots are expected to move towards the equator \citep{starspot}. These values can be seen in Table \ref{tab:priors}.

\begin{table}
    \centering
    \begin{tabular}{llr}
        \toprule
        Parameter & Y23 & This Work \\
        \midrule
        Fixed & & \\
        \midrule
        $P_\text{orb}$ (days) & 0.1617006798 & 0.1617006798\\
        $\omega$ & — & 1 \\
        $\beta$ & 0.08 & 0.08 \\
        \midrule
        Fitted & & \\
        \midrule
        $x_1$ (lt-s) & — & [0.219253, 0.219263] \\
        $i$ & [$50^{\circ}$, $90^{\circ}$] & [$\cos{(83.4^{\circ})}$, $\cos{(0^{\circ})}$] \\
        $K_2$ (km/s) & — & $423 \pm 24$ \\
        $q$ & [0.04, 0.12] & — \\
        $T_\text{base}$ (K) & [3500,7000] & $6400 \pm 250$ \\
        $T_\text{irr}$ (K) & — & [0, 5000] \\
        $f$ & — & [0.0, 0.99] \\
        $D$ (kpc) & [1.664, 7.766]& [1, 8] \\
        $A_v$ (mag) & [0.7, 0.9] & $0.67 \pm 0.07$\\
        \midrule
        Star Spot & & \\
        \midrule
        $T_\text{spot, NTT}$ (K) & — & [$-$6000, 0]\\
        $T_\text{spot, NOT}$ (K) & — & [0, 3000]\\
        $R_\text{spot}$ (deg) & — & [0,360]\\
        $\theta_{\text{spot}}$ (deg) & — & $90 \pm 10$ \\
        $\phi_\text{spot, NTT}$ (deg) & — & 36\\
        $\phi_\text{spot, NOT}$ (deg) & — & 90\\
        \bottomrule
    \end{tabular}
    \caption{Parameter constraints placed on the models during the MCMC sampling. Uniform distributions are denoted by  [min, max] while Gaussian priors are noted with mean $\pm$ sigma. The co-rotation factor is $\omega$ and the gravity darkening coefficient is $\beta$. The semimajor axis of the pulsar, $x_1$ is in light seconds. The mass ratio is denoted as $q$ and the filling factor, $r_\text{nose}/r_{L1}$, is denoted as $f$. The Y23 mass ratio definition is $M_\text{C}/M_\text{PSR}$, whereas we use $M_\text{PSR}/M_\text{C}$.}
    \label{tab:priors}
\end{table}

\subsection{MCMC Sampling}\label{MCMC}
We use the \textsc{ICARUS} light curve models and our MCMC sampling code to find the best-fit values for the free parameters that are required to create the models. For this purpose, we use the \textsc{EMCEE} python package \citep{FM2019} with an ensemble sampler to explore the multi-dimensional parameter space with 20 walkers, that we allow to move for a chain of $10^5$ steps. These parameters include $x_1$ (lt-s), $i$ (rad), $K_2$ (m $\text{s}^{-1}$), $f$, $T_\text{base}$ (K), $T_\text{irr}$ (K), m-M, and $A_V$ (mag). We convert from $x_1$ to $q$ as explained in Section \ref{icarus}.
In our log likelihood function, we create the \textsc{ICARUS} model from the inputted parameters, and are returned a $\chi^2$ for the model given the dataset from \textsc{ICARUS}. We convert this $\chi^2$ into a log likelihood according to:

\begin{equation}
     \log({L}) = -\frac{1}{2} \times \chi^2
\end{equation}
where $L$ is the likelihood. We add the log likelihood with the log prior to get the log probability, which the sampler uses to determine the posterior distribution.

For all of the parameters with a Gaussian distributions with a mean and $1\sigma$ estimates, we initialize our walkers in a sphere of values that are centered at the mean values and have a radius equal to the $1\sigma$ values. For the parameters following a uniform distribution, we initialize our walkers in a sphere that is centered in the middle of our prior range. The radius for $x_1$ is the reported uncertainty in \cite{Sanpa16}. The radius for $T_\text{irr}$ is $100$\,K, $0.1$ for $f$, $0.5$ for m-M. For $i$, we center the sphere at $60^\circ$ with a radius of $10^\circ$. When we initialize our hot spot parameters, we hold $\phi_\text{spot}$ constant, but initialize $\theta_\text{spot}$ like we do with direct heating Gaussian parameters. We also initialize our walkers for $T_\text{spot}$ and $R_\text{spot}$ at the values in the center of their uniform prior ranges. The radius for $T_\text{spot}$ is $100$\,K, and for $R_\text{spot}$ is $10^\circ$.

We check for convergence with the auto-correlation time and run our chains 3000 times longer than the auto-correlation time for each parameter, after excluding a burn in of $60,000$ steps. We also check our results with a different (nested) sampler, \textsc{DYNESTY} \citep{dynestycode, dynestycode2, dynestynested, dynestynested2, dynestymultiellip}, and obtain consistent results for all of our models. 

\subsection{Linked MCMC Runs}
We implement a novel MCMC sampling algorithm that searches the entire parameter space for both the NTT and NOT datasets simultaneously, while forcing most of the parameters to be the same for each step in the chain of the ensemble sampler. Simultaneous fitting of multiple datasets is common when analyzing X-ray data \citep{xspec, isis} and has been implemented for optical data of other types of binary systems \citep{cvfit}, but it has not been implemented yet for optical data of spider systems to the best of our knowledge. The parameters that are linked are those that are not expected to change on short timescales: $x$, $i$, $K_2$, $f$, m-M, and $A_V$. The parameters which are allowed to vary independently in the fits of both datasets are $T_\text{base}$, $T_\text{irr}$, $T_\text{spot}$, $R_\text{spot}$, and $\theta_{\text{spot}}$. We 
fix the azimuthal angle of the hot spot to two different values
%have two $\phi_{\text{spot}}$ parameters 
in our linked fit to the NTT and NOT datasets, 
$\phi_{\text{spot, NTT}}$=36$^\circ$ and $\phi_{\text{spot, NOT}}$=90$^\circ$, respectively.
%
%, both of which are held fixed to improve computational efficiency by reducing the number of free parameters. 
%
%By looking at the , 
%
We choose these values to match the residuals of the symmetric model fits.
%we find the values to set $\phi_{\text{spot}}$ for each dataset. 
%
With this, we extend our parameter space to be 16 dimensional, where our model $M$ is comprised of the model that we fit to the NTT data $M_\text{NTT}$ and the model that we fit to the NOT data $M_\text{NOT}$:

\begin{equation}
    M = M_\text{NTT} + M_\text{NOT} \text{.}
\end{equation}
Each model $M_i$ has 11 free parameters \{$x$, $i$, $K_2$, $f$, $T_\text{base,i}$, $T_\text{irr, i}$, m-M, $A_V$, $T_\text{spot, i}$, $R_\text{spot, i}$, $\theta_\text{spot, i}$\}. Therefore, the full set of parameters of $M$ are \{$x$, $i$, $K_2$, $f$, $T_\text{base,NOT}$, $T_\text{base, NTT}$, $T_\text{irr,NOT}$, $T_\text{irr, NTT}$, m-M, $A_V$, $T_\text{spot, NTT}$, $T_\text{spot, NOT}$, $R_\text{spot, NTT}$, $R_\text{spot, NOT}$, $\theta_\text{spot, NTT}$, $\theta_\text{spot, NOT}$\}. Since the datasets are independent of each other, the $\chi^2$ returned by \textsc{ICARUS} for each $M_i$ is independent of the other, allowing us to sum the $\chi^2$ statistic for each model to give us the $\chi^2$ for $M$ \citep{chi2, chi2gen}:

\begin{equation}
    \chi_\text{M}^2 = \chi_\text{NTT}^2 + \chi_\text{NOT}^2 \text{.}
\end{equation}
From this, the reduced $\chi^2$ is as follows:

\begin{equation}
    \chi_\text{M,dof}^2 = \frac{\chi_\text{M}^2}{n_\text{NTT}+n_\text{NOT}-p_\text{NTT}-p_\text{NOT}-P}
\end{equation}
where $n_\text{dataset}$ is the number of data points in the dataset, $p_\text{dataset}$ is the number of parameters that are independent to that dataset, and P is the number of shared parameters.

\section{Results}\label{results}
We use our MCMC code for both independent and linked fitting of J1622 with our independent temperature constraints from the optical spectroscopy combined with our photometry from NTT in 2017 and NOT in 2022 to obtain our estimates for the parameters of this system.

\subsection{Optimal Subtraction}\label{optsub}

The optical spectra of J1622 (Figure \ref{fig:optsub_spectra}, middle) show clear hydrogen
Balmer absorption lines ($\beta \text{, } \gamma \text{, and } \delta$) as well
as MgI triplet at 5167/73/84\,\AA, all typical of F spectral types.
We considered in this analysis the $4000-5750\,\text{\AA}$ spectral range,
which also includes blends of narrow and fainter metallic lines (many
of which can be seen in Figure \ref{fig:optsub_spectra}).
We find that an F6V star provides the best match to our J1622 spectra,
scaled by a factor $f$=0.77$\pm$0.02.
% (which implies that about 77\% of the flux in this band comes from
%the companion).
The F6V template, average J1622 spectrum and residuals of the optimal
subtraction are shown in Figure \ref{fig:optsub_spectra}.
We note that the F6V template shows a clear CH absorption band around
4300~\AA, seen in spectral types later than F3, which we do not
detect in J1622.
Instead, the MgI triplet lines in J1622 are somewhat stronger than our
best-matching F6 template, as can be seen upon close inspection of the
residuals (Fig. \ref{fig:optsub_spectra}, top).
Together, this suggests that C and Mg in J1622 may be under- and
over-abundant, respectively, as compared to Solar metallicity standard
stars of the same $T_\text{eff}$.

\begin{figure}
    \centering
    \includegraphics[trim={2cm 2cm 1.16cm 0cm}, clip, scale=0.33]{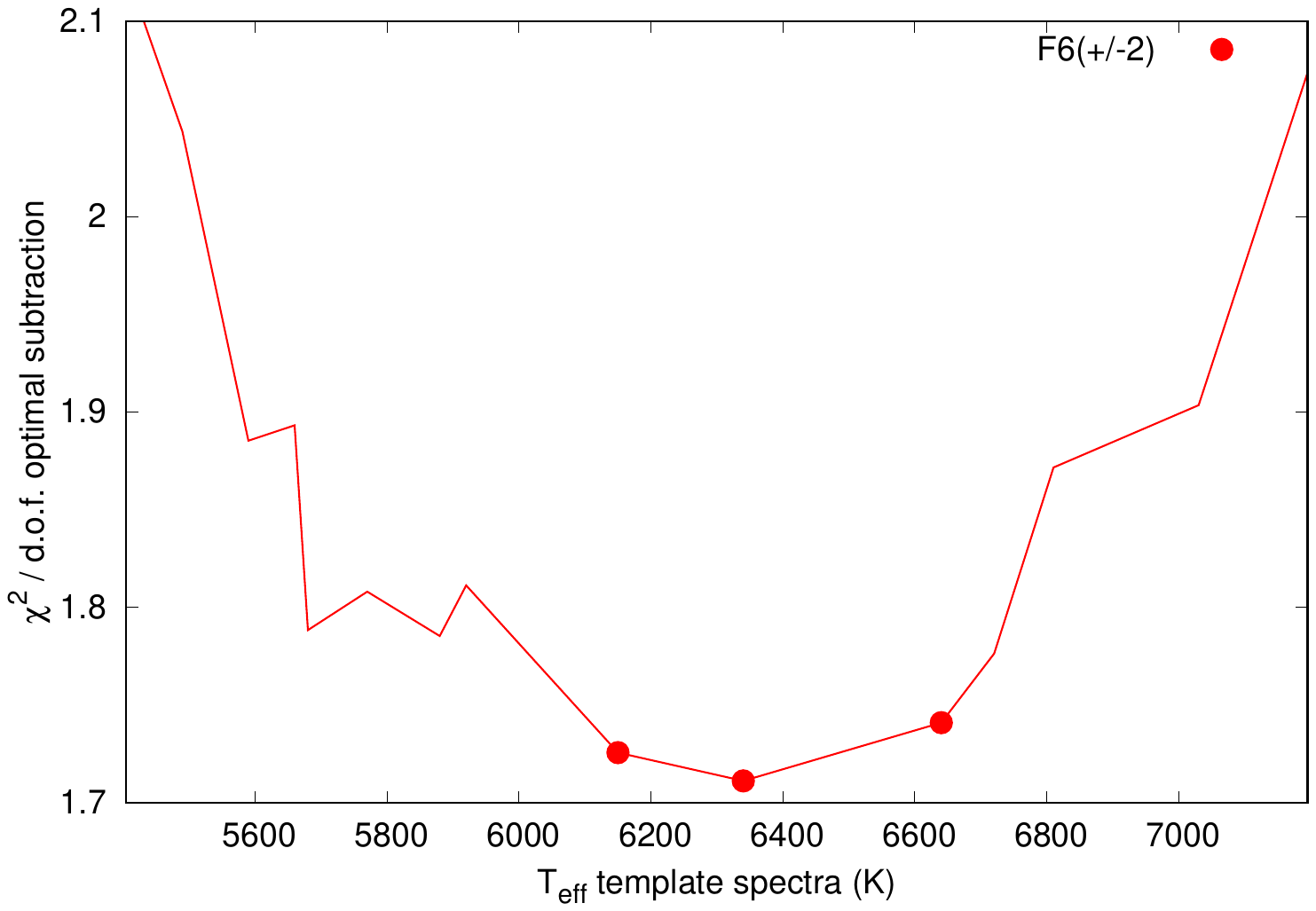}
    \caption{Reduced $\chi^2$ from optimal subtraction vs. template effective temperature (T$_\text{eff}$). The accepted range of T$_\text{eff}$=6400$\pm$250~K is shown with red circles.}
    \label{fig:optsub_chi}
\end{figure}

To quantify the temperature constraints given in Section \ref{soar}, the scatter in the optimal
subtraction residuals is measured and minimized, by computing a
chi-squared between those residuals before and after applying a
Gaussian smoothing.
The results are shown in Figure \ref{fig:optsub_chi}, where the minimum reduced
chi-squared corresponds to the best-matching $T_\text{eff}$=6340~K (F6V)
template.
We estimate a 99\% statistical uncertainty $\simeq$100~K for $\Delta \chi^2$=9.2 (2 parameters, $f$ and $T_\text{eff}$).
Differences in metallicity between the template reference stars and
J1622 can introduce an additional systematic uncertainty in the
temperature determination.
Upon inspection of the residuals we find that spectral types between
F4 and F8 give acceptable results; this range covers the CH and MgI
line intensities mentioned above.
Thus, assuming a systematic spectral type uncertainty of $\pm$2, we
find that our final temperature constraint is $T_\text{eff}$=6400$\pm$250~K.

\subsection{Symmetric Model Fits}
\label{sec:symmetric}

\begin{figure*}
    \subfloat{
        \includegraphics[trim={0cm 0cm 1.15cm 0cm}, clip, scale=0.38]{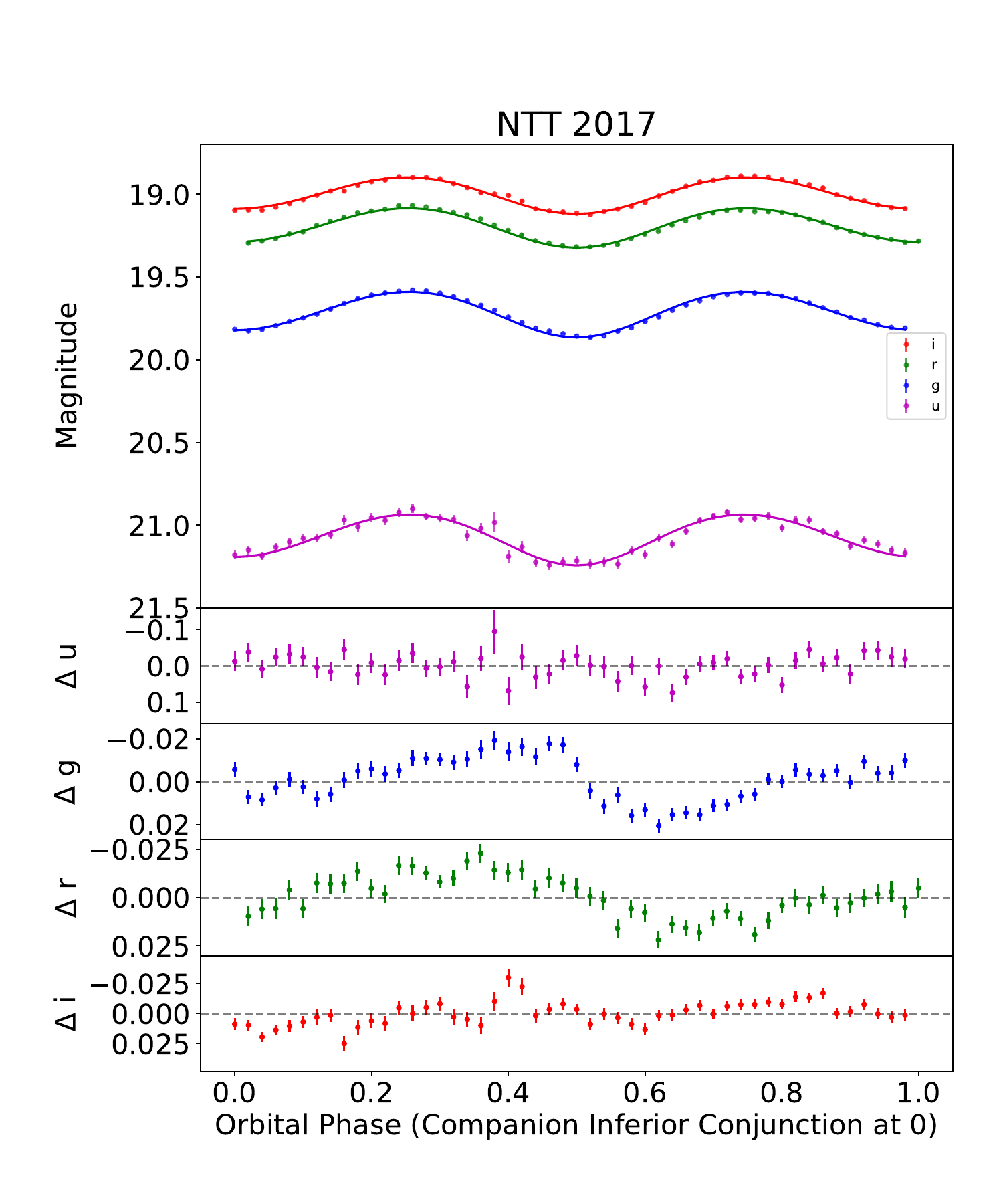}
        }
    \subfloat{
        \includegraphics[trim={1.03cm 0cm 0cm 0cm}, clip,scale=0.38]{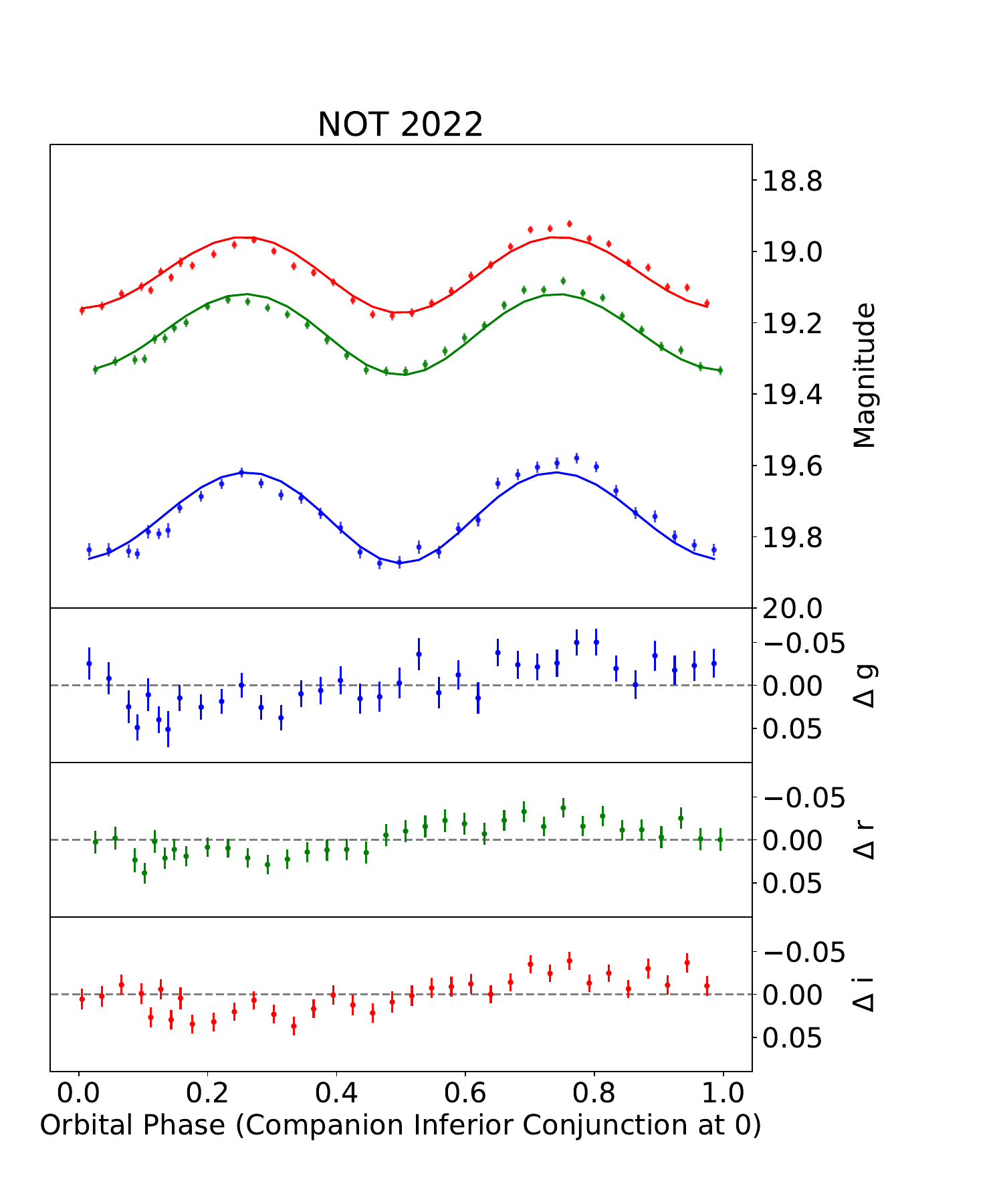}
        }
    \caption{Best-fit symmetric model multiband light curves of PSR1622-0315 from NTT 2017 (left) and NOT 2022 (right). Main panels on top show model light curves (solid line) and data points, while subpanels below show the residuals, with a dashed line at 0 to show how the residuals deviate more clearly}.
    \label{fig:symfitmag}
\end{figure*}

We first fit each of the separate datasets with symmetric, direct heating models, which model outgoing flux from the companion with the effects of gravity darkening and irradiation. We do not make any assumptions on $T_\text{irr}$ and allow it to be a free parameter. The resulting best-fit parameters are reported in Table \ref{tab:derived}. 
% We find that with a symmetric direct heating model including irradiation, the NTT dataset gives a high pulsar mass of $2.61_{-0.39}^{+0.45} M_{\odot}$, while the NOT dataset gives a high, but more modest pulsar mass of $2.16_{-0.56}^{+0.63} M_{\odot}$. The NTT data also gives a higher companion mass estimate than the NOT data by $0.02 M_{\odot}$.
We note that the best-fit NTT $K_2$ is higher than the \cite{Strader19} measurement of $423 \pm 8\,\text{ km s}^{-1}$ by more than $1\sigma$. The best-fit base temperature of $6793$\,K is also over $1 \sigma$ from the results from optimal subtraction. This high base temperature leads to a high best-fit extinction of $0.81 \pm 0.06$. 

In comparison, the NOT best-fit $K_2$ and base temperature are well within $1 \sigma$ of our priors from Table \ref{tab:priors} as well as the \cite{Strader19} reported $K_2$ and spectroscopic measurement of the base temperature. Additionally, the best-fit extinction value also closely follows its prior. When comparing these two fits, we also find that the irradiation temperature for the NOT fit is greater than the irradiation temperature of the NTT fit by $1160$\,K, leading to an irradiating luminosity that is one order of magnitude greater than that of the NTT fit. The $\chi^2_{\rm dof}$, of both symmetric fits are high, greater than $3$.

To test these models, we also applied a wide and flat prior on $T_\text{base}$ between 1000 and 10000\,K, instead of the Gaussian prior described in Section~\ref{priors}. The corresponding best-fit solutions had $T_\text{base}$ much larger (8000-9000\,K) than what is required by the spectroscopic observations (Sec.~\ref{optsub}). Therefore, we find that the spectroscopic line-based prior on $T_\text{base}$ is needed to constrain the temperature of the model, for both the NTT and NOT data. 
%We have applied this prior on $T_\text{base}$, which is the temperature of the star before the effects of gravity darkening, irradiation, and star spots are added to the model. 
While the spectroscopic measurements of the effective temperature depend on the orbital phase and inclination of the system during the observations, we notice that due to the low irradiation, the difference between the day and night side temperatures is minimal and therefore, the effect of the orbital phase of the spectroscopic observations is negligible. Indeed, as mentioned in Section~\ref{soar}, the spectral type did not significantly change after averaging two and ten spectra. When we calculate the hemisphere averaged $T_\text{day}$ and $T_\text{night}$, which are closer to the temperature measured from spectroscopy, we find that these values are within $250\,K$, or $1\sigma$, of the prior value of $6400K$. Therefore, applying the temperature prior on $T_\text{base}$ for J1622 leads to results that are consistent with those that would have been found when applying the prior on $T_\text{day}$ or $T_\text{night}$.

We also applied our MCMC fitting routine after relaxing our band calibration error of 0.01\,mag to 0.05\,mag for both datasets. We find that doing so returns a best fit model with a higher $T_\text{base}$ than is expected from optimal subtraction, for only the NTT data. This is likely due to the fact that most of the band uncertainties are well below 0.01\, mag when fitting the NOT data (Table \ref{tab:priors}). For the NTT data, however, most of the band uncertainties are greater than 0.01 in magnitude. Therefore, changing the allowed band calibration error affects the NTT fits as higher temperature solutions are allowed with the 0.05\,mag error but ruled out with the smaller error of 0.01\,mag.

The model light curves from our symmetric fits are shown Figure \ref{fig:symfitmag}. For both best-fit model light curves, we find clear trends in the \textit{g'}, \textit{r'}, and \textit{i'} band residuals. With the NTT dataset, the symmetric model fits show an underprediction of flux starting around $\phi_\text{orb} = 0.3$ until $0.5$ and an overprediction of flux starting at $\phi_\text{orb} = 0.5$ until $0.8$. In relation to the surface of the companion star, this results in an underprediction of flux on the trailing side of the companion and an overprediction on the leading side. The NOT fits show the opposite trend: they underpredict the flux at phase $0.75$, which is when the leading side of the companion is seen most clearly. These trends in the residuals reflect the different asymmetry in the maxima of both light curves, which points to the detection of flux from features beyond direct heating and irradiation from the pulsar wind. We explore this by introducing star spots into our models.

\subsection{Star Spot Model Fits}

\begin{figure*}
    \subfloat{
        \includegraphics[trim={0.3cm 0cm 0.26cm 0cm}, clip, width = 0.49\linewidth]{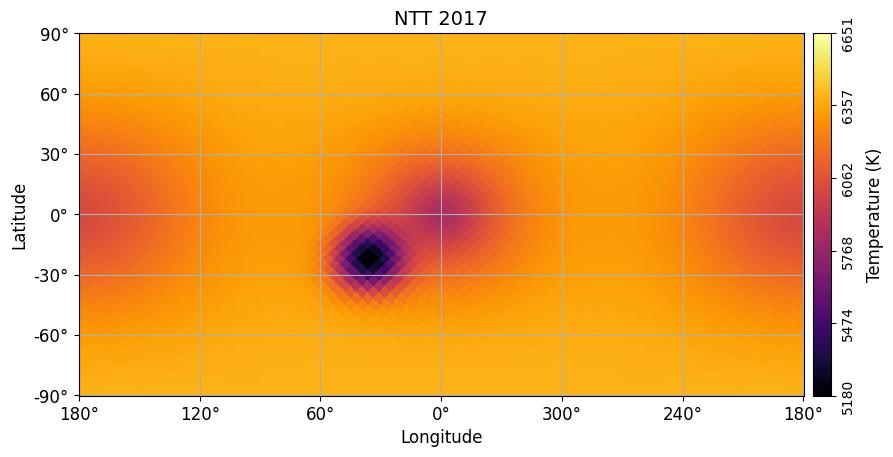}
        }
    \subfloat{
        \includegraphics[trim={0.3cm 0cm 0cm 0cm}, clip, width = 0.49\linewidth]{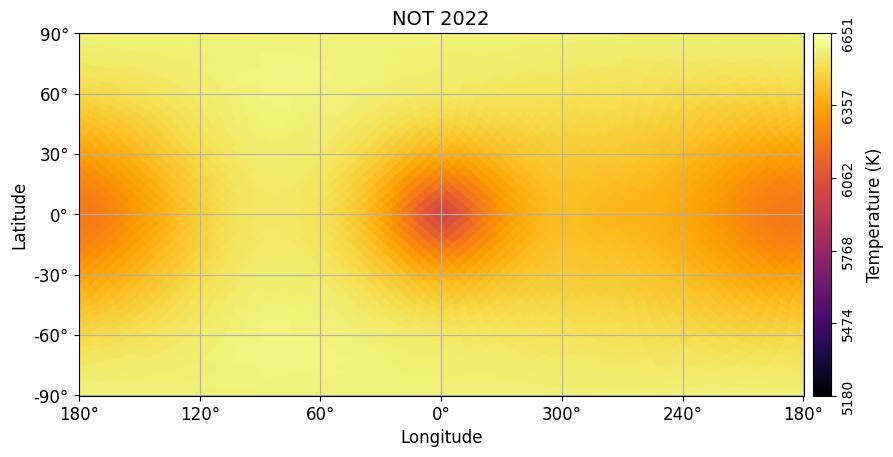}
        }
    \caption{Surface plots of the companion star. The x-axis shows the longitude, or $\phi_\text{spot}$, while the y-axis shows the latitude in terms of degrees from the equator. The NTT model cold spot (left) is best seen at orbital phase 0.6 (corresponding to a longitude of 36$^\circ$) while the NOT hot spot (right) is best seen at orbital phase 0.75 (corresponding to a longitude of 90$^\circ$).}
    \label{fig:surface}
\end{figure*}

 \begin{figure*}
     \subfloat{
         \includegraphics[trim={0cm 0cm 1.15cm 0cm}, clip, scale=0.4]{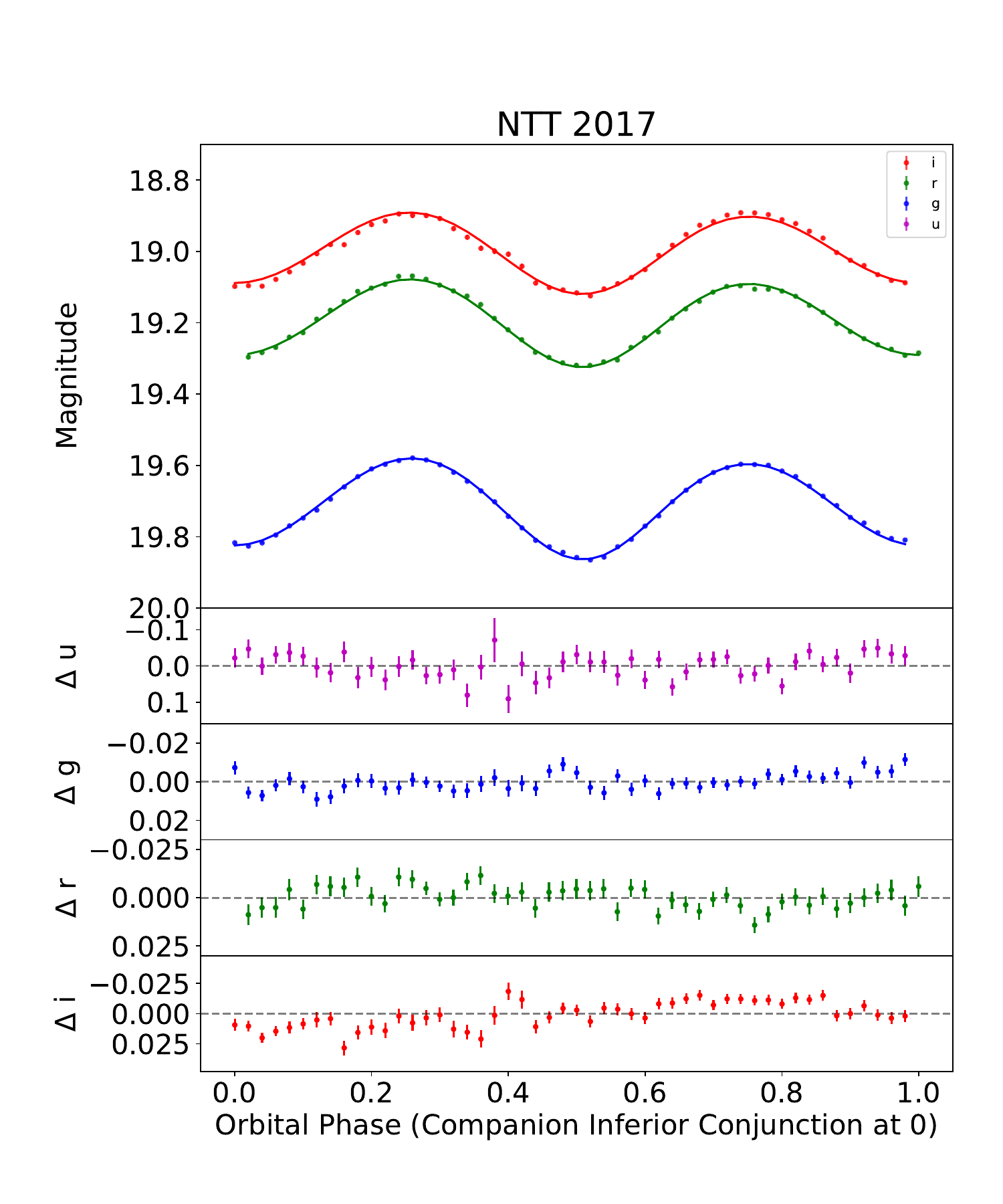}
         }
     \subfloat{
         \includegraphics[trim={1cm 0cm 0cm 0cm}, clip,scale=0.4]{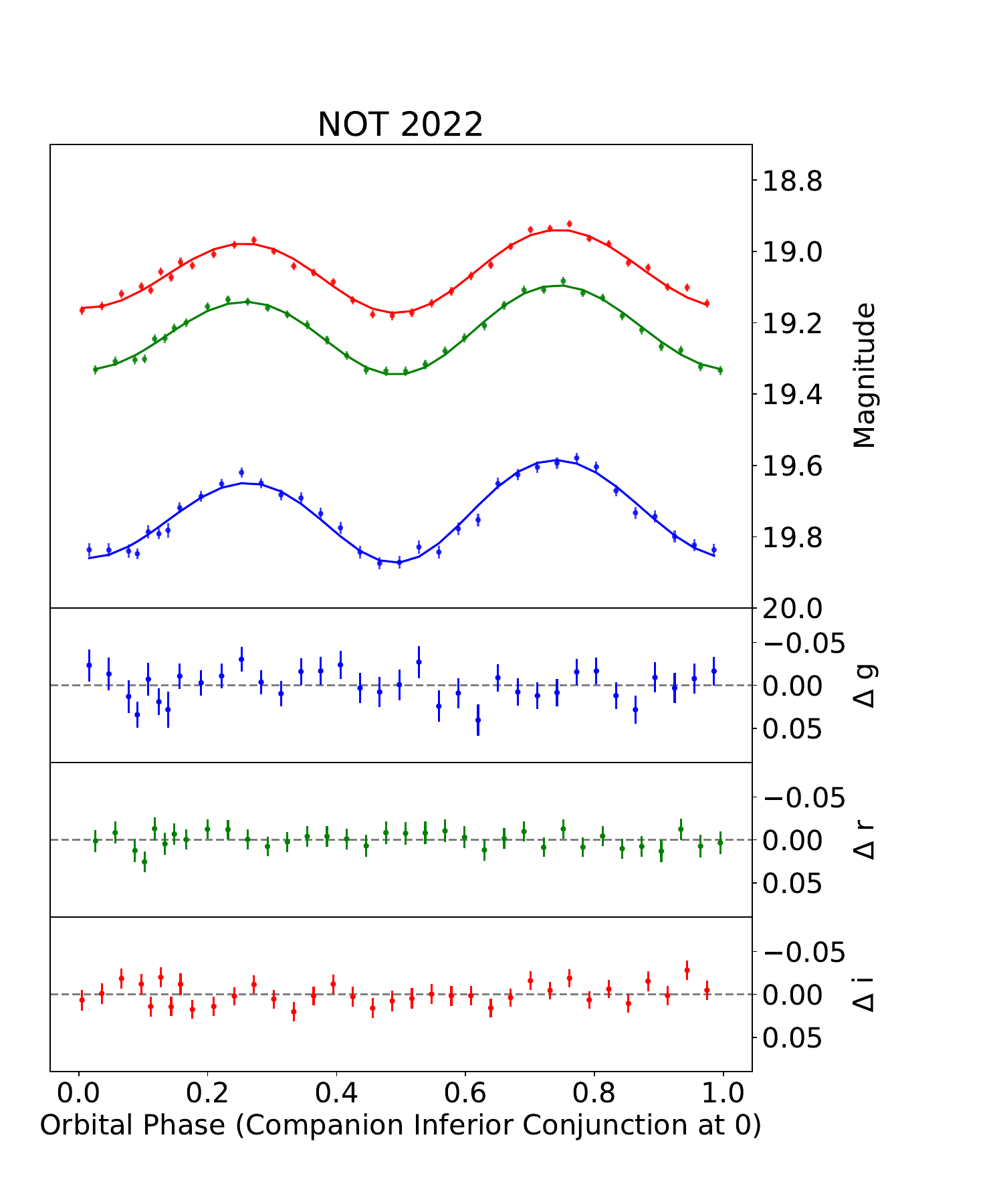}
         }    
     \caption{Best-fit asymmetric model gri light curves of PSR1622-0315 from NTT 2017 (left) and NOT 2022 (right). Main panels on top show model light curves (solid line) and data points, while subpanels below show the residuals, with a dashed line at 0 to show how the residuals deviate more clearly. The NTT best-fit model has a cold spot that can be seen at phase of 0.6 and the NOT has a hot spot that can be seen at phase 0.75.}
     \label{fig:asymfitmag}
 \end{figure*}

To fit the asymmetries that we observe in the light curves and their residuals, we add star spots to our model and then perform our MCMC sampling. We apply a cold spot at $\phi_\text{orbit} = 0.6$ when fitting the NTT data and a hot spot at $\phi_\text{orbit} = 0.75$ when fitting the NOT data. This is the only parameter that we fix in the model. The temperature of the spot, the spread of the spot, and the co-latitude location of the spot are free parameters. The surface of the companion star with the best-fit star spot for each dataset is shown in Fig.\ref{fig:surface}. We find that the total spread of the NTT cold spot is less than 1\% of the entire surface of the companion star, but the NOT hot spot spreads over the entire leading side of the companion and covers over 50\% of the entire surface of the companion star. The nose of the companion is at latitude and longitude of $0^\circ$, and is colder than the surface elements seen at quadrature as well as the night side due to gravity darkening dominating over irradiation. 

The best-fit parameters from these asymmetric models are reported in Table \ref{tab:derived} and the corner plots for the fits are in with corner plots shown in Figures \ref{fig:ucamcscorner} and \ref{fig:nothscorner}. We find lower $\chi^2_{\rm dof}$s than the respective symmetric models, indicating that the star spot models yield significantly improved fits. The NOT hot spot model returns a $\chi^2_{\rm dof}$ that is close to $1$, and while the NTT cold spot model has a $\chi^2_{\rm dof}$ that is above $2$, it has decreased from the symmetric fit value by over $50\%$. With the exception of the NTT \textit{g'} band, the magnitude offsets are smaller for both asymmetric fits than their respective symmetric fit. In addition to this, when we look at the resulting light curves in Fig. \ref{fig:asymfitmag}, we find that the residuals now appear flat in all of the bands.

From these asymmetric models, we find that the NTT dataset gives a pulsar mass of $2.3_{-0.3}^{+0.4}\,\text{M}_{\odot}$, and the NOT dataset gives a slightly lower pulsar mass of $2.1_{-0.5}^{+0.6}\,\text{M}_{\odot}$. These two central values are consistent with each other, and are both lower than the estimates from the direct heating models. After adding star spots, we find that the companion masses from the two fits are also consistent with each other, at $0.14-0.15\,\text{M}_{\odot}$. In addition to the mass estimates, we find that the best-fit values for $i$, K2, f, $T_\text{base}$, distance, and $A_v$ are now consistent between the two datasets. There is still a difference in the irradiation temperature between the two datasets, which also causes the difference in the irradiation luminosity between the two fits. We note that the irradiation temperature of the NTT fit increased from $1900$\,K to $2700$\,K after adding a cold spot, while the irradiation temperature of the NOT fit remains constant at $3000$\,K with an added hot spot. Both of these irradiation temperatures lead to irradiation luminosities that are 2 orders of magnitude lower than that of $\Dot{E}$, which is $7.7 \times 10^{33}\,\text{ erg s}^{-1}$ \citep{Sanpa16}.

\subsection{Linked Fits}

% \begin{figure*}
%     \begin{subfigure}[h]{0.5\linewidth}
%         \includegraphics[trim={0cm 0cm 2.1cm 0cm}, clip, scale=0.4]{240116_ucam_not_toptsub_fixi_symUCAM_LC.pdf}
%     \end{subfigure}
%     \begin{subfigure}[h]{0.5\linewidth}
%         \includegraphics[trim={3.2cm 0cm 0cm 0cm}, clip,scale=0.4]{240116_ucam_not_toptsub_fixi_symNOT_LC.pdf}
%     \end{subfigure}
%     \caption{These are the best-fit symmetric model light curves for the ULTRACAM and NOT datasets in flux densities. These fits were done with linked fitting, where only the base and irradiation temperature of the two datasets were allowed to vary.}
%     \label{fig:symlinkedflux}
% \end{figure*}

\begin{figure*}
    \subfloat{
        \includegraphics[trim={0cm 0cm 1.15cm 0cm}, clip, scale=0.38]{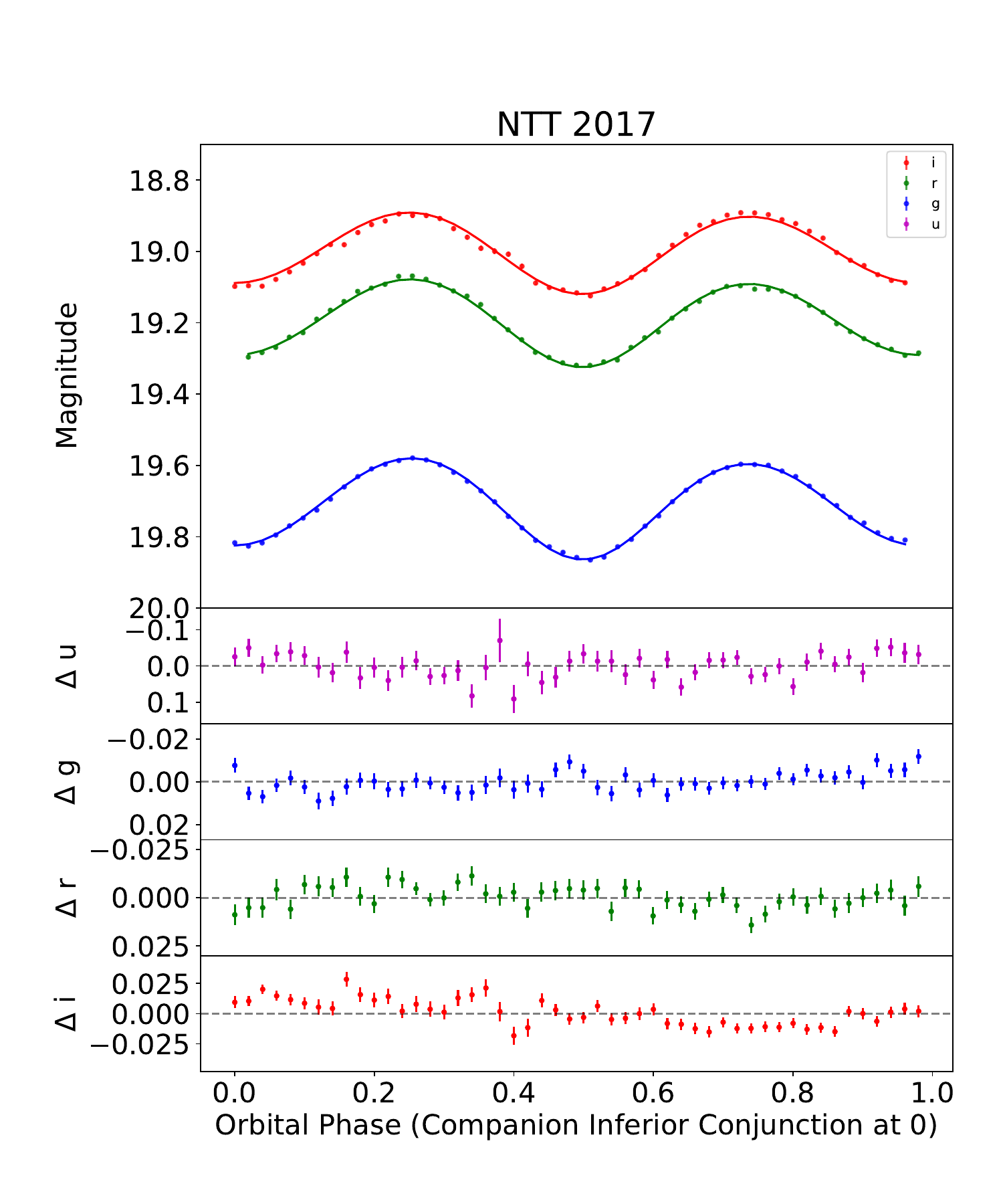}
        }
    \subfloat{
        \includegraphics[trim={1.03cm 0cm 0cm 0cm}, clip, scale=0.38]{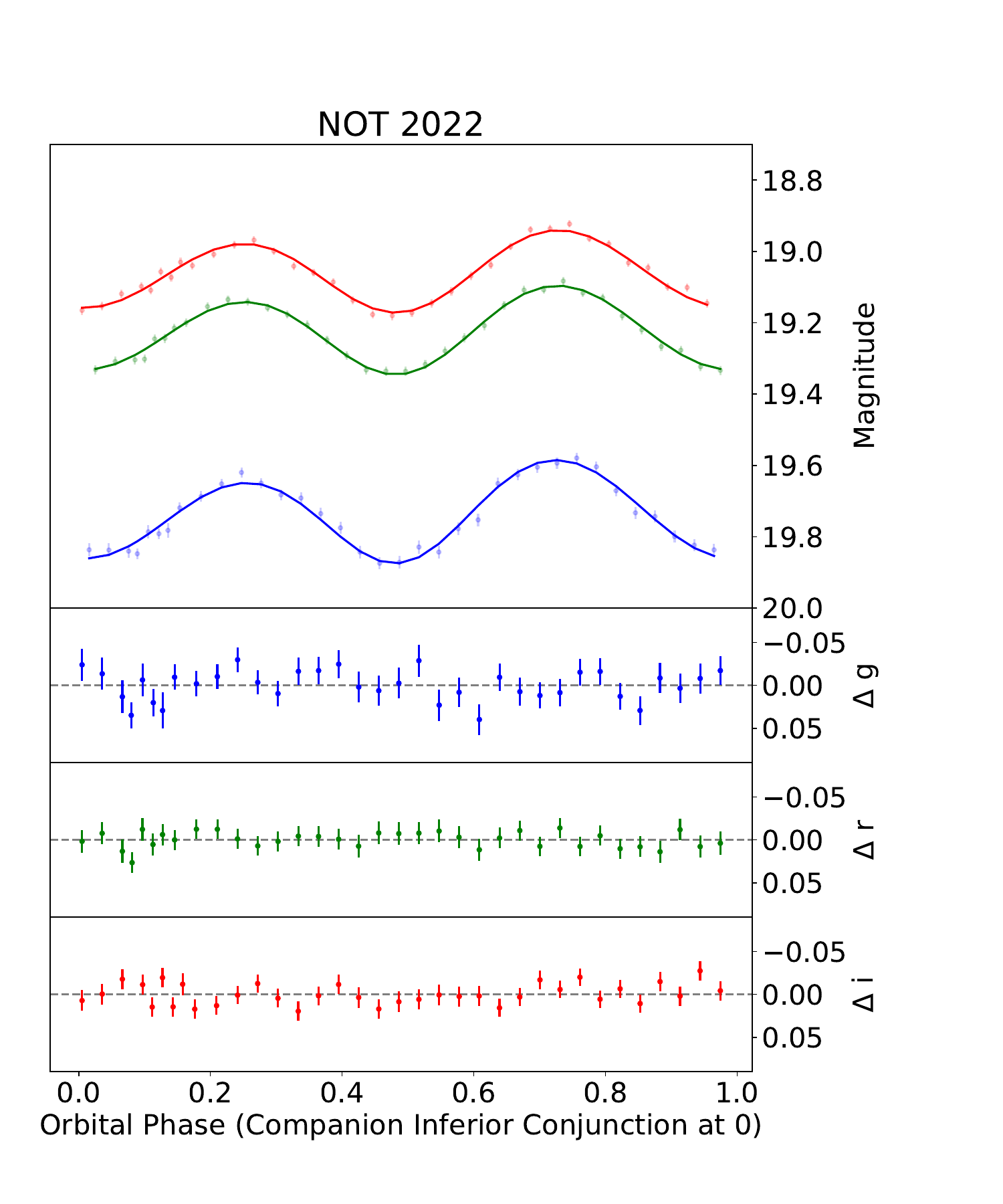}
        }
    \caption{Best-fit asymmetric model light curves and residuals for the linked NTT and NOT fit. Only the base and irradiation temperature of the two datasets, as well as the four star spot parameters, were allowed to vary for this fit. The remaining nine parameters were linked between the two models.}
    \label{fig:linkedmag}
\end{figure*}

\begin{table*}[ht!]
    \centering
%    \begin{tabular}{lrrrrrrr}
    \begin{tabular}{lccccccc}
        \toprule
        Fitted & Y23 & \multicolumn{2}{c}{NTT 2017} & \multicolumn{2}{c}{NOT 2022} & \multicolumn{2}{c}{NTT+NOT}\\
         &  & Symmetric & Asymmetric & Symmetric & Asymmetric & \multicolumn{2}{c}{Asymmetric}\\
        \midrule
        i (deg) & $78.1_{-6.9}^{+7.9}$ & $64.8_{-2.9}^{+2.8}$  & $63.6_{-2.3}^{+2.4}$ & $59.3_{-5.0}^{+12.6}$ & $62.2_{-6.8}^{+12.4}$ & \multicolumn{2}{c}{$62.8_{-2.6}^{+2.4}$} \\
        $K_2$ (km/s) & — & $467\pm22$ & $439\pm22$ & $419\pm24$ & $421\pm24$ & \multicolumn{2}{c}{$437\pm22$} \\
        f & 1 & $0.82\pm0.02$ & $0.83\pm0.02$ & $0.88\pm0.09$ & $0.86_{-0.07}^{+0.10}$ & \multicolumn{2}{c}{$0.84\pm0.02$} \\
        $T_\text{base}$ (K) & $6383_{-98}^{+89}$\footnote{Y23 report $T_\text{eff}$, which accounts for gravity darkening.} & $6793_{-130}^{+160}$ & $6395_{-90}^{+80}$ & $6508_{-100}^{+110}$ & $6190_{-240}^{+210}$ & $6264_{-60}^{+100}$ & $6057_{-140}^{+110}$ \\
        $T_\text{irr}$ (K) & — & $1900_{-160}^{+150}$ & $2676 \pm 70$& $3059 \pm 220$ & $3023_{-210}^{+230}$ & $2626_{-60}^{+70}$ & $2738_{-120}^{+110}$ \\
        $T_\text{spot}$ (K) & — & — & $-1096_{-370}^{+280}$ & — & $434_{-200}^{+250}$ & $-1047_{-350}^{+260}$ & $208_{-70}^{+140}$ \\
        $R_\text{spot}$ (deg) & — & — & $10\pm2$ & — & $190_{-70}^{+62}$ & $10\pm2$ & $127_{-42}^{+50}$ \\ %\begin{tabular}{@{}c@{}}$10\pm2$\\$127_{-42}^{+50}$ \end{tabular}
        $\theta_\text{spot}$ (deg) & — & — & $112\pm5$ & — & $90\pm11$ & $112_{-4}^{+5}$ & $90\pm11$ \\
        D (kpc) & $2.06\pm0.04$ & $2.41\pm0.14$ & $2.14\pm0.10$ & $2.37_{-0.33}^{+0.23}$ & $2.32_{-0.28}^{+0.27}$ & \multicolumn{2}{c}{$2.19\pm0.10$} \\
        $A_v (\text{mag})$ & $0.81_{-0.07}^{+0.06}$ & $0.81\pm0.06$ & $0.71_{-0.08}^{+0.07}$ & $0.66\pm0.06$ & $0.69\pm0.07$ & \multicolumn{2}{c}{$0.56_{-0.05}^{+0.08}$} \\
        \midrule
        Derived &  &  &  &  &  & \\
        \midrule
        q & $15.15_{-1.07}^{+0.72}$ & $15.8 \pm 0.7$ & $14.9_{-0.7}^{+0.8}$ & $14.2 \pm 0.8$ & $14.2 \pm 0.8$ & \multicolumn{2}{c}{$14.8_{-0.7}^{+0.8}$}\\
        $M_1 (M_{\odot})$ & $1.84\pm0.19$ & $2.6_{-0.4}^{+0.5}$ & $2.3_{-0.3}^{+0.4}$ & $2.2 \pm 0.6$ & $2.1_{-0.5}^{+0.6}$ & \multicolumn{2}{c}{$2.3 \pm 0.4$}\\
        $M_2 (M_{\odot})$ & $0.122_{-0.006}^{+0.007}$ & $0.17\pm0.02$ & $0.15\pm0.02$ & $0.15\pm0.04$ & $0.14_{-0.03}^{+0.04}$ & \multicolumn{2}{c}{$0.15\pm0.02$}\\
        $K_1$ (km/s) & — & $29.6\pm0.0$ & $29.6\pm0.0$ & $29.6\pm0.0$ & $29.6\pm0.0$ & \multicolumn{2}{c}{$29.6\pm0.0$} \\
        $T_\text{day}$ (K) & — & $6598_{-130}^{+160}$ & $6211_{-90}^{+80}$ & $6353_{-100}^{+110}$ & $6428_{-250}^{+220}$ & $6085_{-60}^{+100}$ & $6071_{-140}^{+110}$ \\
        $T_\text{night}$ (K) & — & $6609_{-130}^{+160}$ & $6222_{-90}^{+80}$ & $6330_{-100}^{+110}$ & $6404_{-250}^{+220}$ & $6095_{-60}^{+100}$ & $6052_{-140}^{+110}$ \\
        $L_\text{irr}$ ($10^{31}$ erg/s) & — & $0.4\pm0.05$ & $1.5\pm0.6$ & $2.7\pm0.1$ & $2.4\pm0.1$ & $1.4\pm0.6$ & $1.7\pm0.7$ \\ 
        \midrule
        Model Fit &  &  &  &  &  & \\
        \midrule
        u Offset (mag) & — & 0.099 & 0.022 & — & — & 0.024 & — \\
        g Offset (mag) & — & -0.025 & -0.037 & 0.012 & 0.004 & -0.032 & -0.029 \\
        r Offset (mag) & — & -0.034 & 0.012 & 0.001 & -0.001 & 0.002 & 0.008 \\
        i Offset (mag) & — & -0.060 & 0.004 & 0.014 & 0.004 & -0.020 & -0.024 \\
        $\chi^2_\text{dof}$ & — & 6.00 & 2.64 & 3.33 & 1.17 & \multicolumn{2}{c}{2.08} \\
        \bottomrule
    \end{tabular}
    \caption{Fitted and derived parameters from the MCMC fitting, with the 50th percentile value reported with the 16 and 84 percentiles as uncertainties. The model on the far left is that from \cite{Y23}. The next two columns are from fitting the NTT dataset with symmetric and cold spot models. The negative temperature of the star spot indicates a cold spot. The following two columns are from fitting the NOT dataset with symmetric and hot spot models. For parameters that are independent in the linked fit, the NTT value is given first and then the NOT value is on the right. Corner plots of the fit results are shown in Appendix \ref{corners}.}
    \label{tab:derived}  
\end{table*}

We apply our linked fitting routine to find the best-fit asymmetric models for the NTT and NOT datasets, where the base temperature, irradiation temperature, and star spot parameters are independent between the two models, and the rest of the parameters are fixed for both datasets. The resulting fit is reported in Table \ref{tab:derived}, with corner plots shown in Fig. \ref{fig:linkedcorner}. This fit is the best model out of the five models we explored, with a $\chi^2_{\rm dof}$ of $2.08$. This is a lower value than that of the NTT individual fit, indicating an improvement over the unlinked NTT fit. While the NOT individual asymmetric fit has a better $\chi^2_{\rm dof}$, we note that most of the parameter values agree between the two fits. 

In fact, all of the linked fit parameter values are consistent with both individual asymmetric fits, with the exception of $A_v$. The derived parameters $q$, $M_1$, and $M_2$ also agree with the previous independent asymmetric fits. We again find a high neutron star mass of $2.3 \pm 0.4\, \text{M}_\odot$, which is consistent with the estimates from the independent symmetric and asymmetric model fits. Because of the high level of agreement between the linked and individual models, we can see that the light curves from the linked fitting in Fig. \ref{fig:linkedmag} appear similar to the ones from the independent fits.
% The surface plots of these model companion stars also look identical to the ones in Fig. \ref{fig:surface}. 

We find a higher base temperature for the NTT linked fit compared to the NOT linked fit, but the opposite trend for the irradiation temperature. The NTT and NOT linked irradiation temperatures are within the NOT linked irradiation temperature uncertainty, but slightly below $2\sigma$ of the NTT uncertainty. The irradiation temperature for the linked NTT model differs from the independent NTT model by less than $1\sigma$, whereas the irradiation temperature for the NOT linked and independent models differ by less than $1.5\sigma$. This difference in the NOT independent and linked fit irradiation temperature leads to the difference in $L_\text{irr}$ between these two models, which are again 2 orders of magnitude below $\Dot{E}$. With the exception of the NTT cold spot temperature and the NOT hot spot radius and temperature, the star spot parameters are also the same between the linked and individual asymmetric fits.

The differences between the linked and independent NTT asymmetric fits are in the base and irradiation temperatures, as well as extinction. The lower base and irradiation temperatures and lower extinction better fit the NTT data. The \textit{u'} and \textit{i'} band offsets are smaller for the independent NTT fit, while the \textit{g'} and \textit{r'} band offsets are smaller for the linked fit. For the NOT fits, the differences are in the irradiation temperatures, hot spot temperature, and extinction, and the independent fit parameters slightly match the data better. In particular, the higher irradiation and spot temperature, as well as the higher extinction value better fit the NOT data. 

\section{Discussion}\label{discussion}

\subsection{Asymmetric Minima and Low-Level Irradiation}
\label{sec:irradiation}

In all of our fits we find low but nonzero irradiation in the companion star of J1622. The irradiation luminosity required to reproduce the optical light curves is at most $0.3\%$ of $\Dot{E}$, and as low as $0.1\%$ of $\Dot{E}$ (see Table~\ref{tab:derived}). This is in contrast with previous work on this system by Y23, where it was found that the difference in log likelihoods between models with and without irradiation is less than 0.01\% \citep{Y23}. Due to this, Y23's MCMC parameter search was done without including irradiation. When we compare our fits with that of Y23, we do find that T$_\text{base}$ is consistent between our asymmetric fits and Y23, at close to $6400$\,K. However, we find that at these base temperatures, our model light curves have significant residuals around inferior conjunction of the companion when we fix $T_\text{irr} = 0$\,K. This is due to our light curves showing asymmetric minima at superior and inferior conjunction of the companion not observed by Y23, since we have better data with higher signal to noise photometry. We see this in both our symmetric and asymmetric models for effective temperatures that agree with our optimal subtraction results. Therefore, we need to include low, but nonzero, irradiation in our models in order to match the data well.

In \cite{Turchetta23}, it was found that both the (g$-$r) and (r$-$i) colors are constant along the orbit. Our 2017 NTT light curves also have two maxima per orbit, showing that there are no qualitative changes in the irradiation of the companion on a five-year timescale. While we notice that the irradiation luminosities are between $1$ and $2\sigma$ of each other in our independent fits of the two datasets, the two irradiation luminosities differ by less than $1\sigma$ in our linked fit model. Therefore, we find no evidence of a change in the irradiation between these two datasets. However, we find that despite the flat colors observed in this system, we require models with significant irradiation to fit the data well at all orbital phases. The base temperature of J1622 is higher than many redback systems \citep[cf. our Table \ref{tab:derived} and Table 1 of][]{Turchetta23}, but even with these high temperatures, we find that models without irradiation do not fit the asymmetric minima well. Therefore, it is necessary to include the effects of irradiation to properly fit our light curves from J1622. We conclude that careful modelling of high-quality optical light curves can reveal subtle irradiation effects in spider binaries.

\subsection{Variable Asymmetric Maxima: Star Spots or a Variable Shock?}

The optical light curves of J1622 show both asymmetric minima and maxima, and we find that the asymmetry in the maxima has changed over five years (see Figure~\ref{fig:symfitmag} and Section~\ref{sec:symmetric}). In 2017 with NTT, we observed more flux between orbital phase $0.25$ and $0.5$ and therefore a larger maxima here, than  between orbital phase $0.5$ to $0.75$. In 2022 with NOT, we observed more flux at orbital phase $0.75$, which is the opposite side. Variable asymmetry in light curve maxima over timescales of 5-10 years has been observed in other redback systems. PSR J2039-5617 showed light curves with two maxima per orbit, and the magnitude difference between these maxima was  variable over five years \citep{j2039}.
Unlike J1622, however, the second maximum (companion descending node) was consistently dimmer than the first (companion on the ascending node). 
%In the case of PSR J2051-0827, \cite{j2051} observed asymmetrical optical light curves in 2011, and symmetric light curves in 2021.

We find a more extreme change in J1622, with the phase of the absolute maximum changing (from about 0.25 to about 0.75, see Figure~\ref{fig:linkedmag}) over five years. The best-fitting asymmetric models have a cold spot at orbital phase $0.6$ for the 2017 data, and a hot spot at orbital phase $0.75$ for the 2022 data. Both spots are on the leading side, but the temperature difference due to the cold spot is five times greater than that of the hot spot. These differences in fluxes flattens the residuals considerably, showing that the asymmetries were captured in the models better than with the symmetric models (cf. Figures~\ref{fig:symfitmag} and \ref{fig:asymfitmag}). 

While the spread of the cold spot is contained to the side seen at quadrature and covers less than $1\%$ of the whole surface of the companion, we find that the spread of the hot spot wraps around the star, and covers $80\%$ of the star. Since the hot spot covers most of the star, the surface of the companion appears to have a higher $T_\text{base}$ in 2022 than in 2017, even though the model in 2022 gives $200$\,K lower $T_\text{base}$. It is possible that the star spots change due to changes in the magnetic field of the companion, as seen in PSR J1723-2837, where multiple star spots were observed with lifetimes on the order of a few days to 1.5 months \citep{j1723}. One or several additional effects like diffusion, convection, and wind heating \citep{reheat,windreheat} could also explain the asymmetries we observe in this system.

Another possibility is that the difference in light curve asymmetries is due to the presence of an intrabinary shock.
If there is an intrabinary shock, the X-ray and gamma-ray fluxes should show orbital variability. The X-ray emission detected has a luminosity $L_X$ of $4.1\times10^{30}$\,erg/s. There were not enough counts to determine any orbital variability of the system, but the hard X-ray emission is most likely non-thermal and fits a power law of index $2.0 \pm 0.3$ \citep{Gentile2018}. This spectral index is consistent with that of an intrabinary shock as seen in other systems like that of PSR B1957+20, which has a spectral index of $1.9 \pm 0.5$ \citep{x1957}. While \cite{Sanpa16} find that there could be evidence for gamma ray pulsations of J1622, and significant pulsation have been confirmed by \citep{3pc}. 

In the optical wavelength range, models by \cite{romani16} that include intrabinary shock have been used to better fit asymmetric light curves. The pulsar and companion wind parameters determine which object the intrabinary shock wraps around, as well as how the optical light curves change shape due to particles and radiation from the shock. For some particular parameter values for the two winds, there can be a brightening of more than $0.02$ mag at orbital phase $0.25-0.5$ accompanied by a dimming of more than $0.04$ mag at orbital phase $0.5-0.75$ \citep{romani16}. The asymmetric changes we see in our light curves from either year therefore, could be modeled by adding an intrabinary shock with specific parameter values. 
Thus, we suggest an alternative interpretation of the observed variable asymmetry in J1622: an intrabinary shock shape/geometry which is variable on multi-year timescales. This could in turn be due to variability in the wind of the 0.15~M$_{\odot}$ companion star. Modelling of the optical light curves of J1622 using intrabinary shock irradiation, beyond the scope of this work, can test this hypothesis.

\subsection{Support for a Super-Massive Neutron Star}

\begin{figure}
    \centering
    \includegraphics[trim={2cm 1.5cm 0cm 0cm}, clip, scale=0.33]{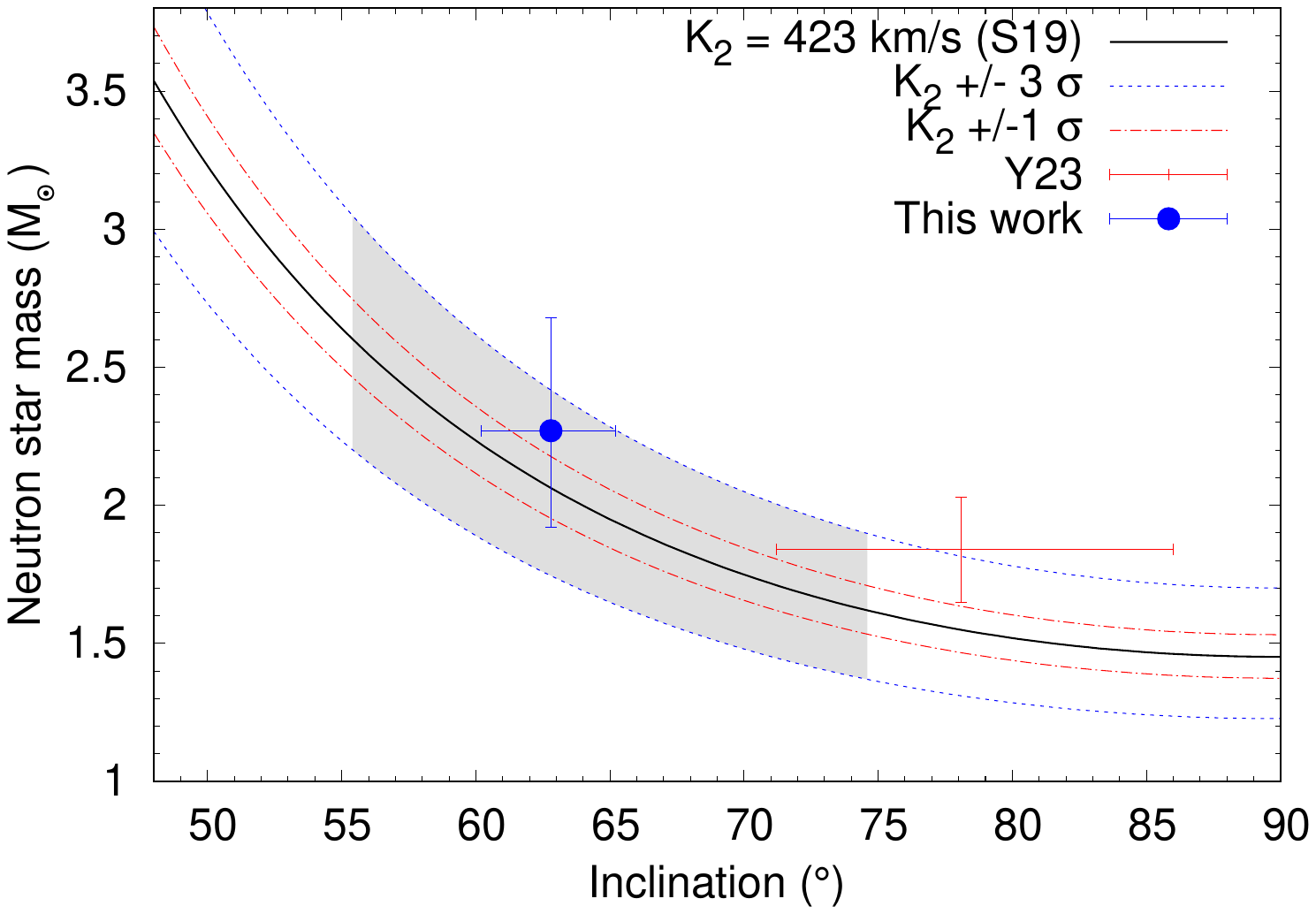}
    \caption{Constraints on $M_\text{NS}$ from $i$ and $K_2$. The central value of $K_2$ is from spectroscopic constraints from \citep{Strader19}, with the $1 \sigma$ level shown with red dashed lines and the $3 \sigma$ level denoted with blue dashed lines. The range of possible $i$ values found from the models used in this work place limits on the gray region, which shows the range of possible $M_\text{NS}$ values consistent with conservative estimates of $K_2$. Y23 derived $M_\text{NS}$ shown in red, while $M_\text{NS}$ from our combined asymmetric fit is shown in blue.}
    \label{fig:K2iM}
\end{figure}

We find a central value for the neutron star mass higher than $2$\,M$_{\odot}$ in all of our fits (Table~\ref{tab:derived}). The NOT asymmetric individual fit gives our lowest estimate of the neutron star mass, at $2.1_{-0.5}^{+0.6} M_\odot$, while the linked and individual asymmetric NTT fits give higher neutron star masses of $2.3 \pm 0.4 M_\odot$. Thus, we find higher neutron star masses than Y23, yet with higher uncertainties (they reported $1.84 \pm 0.19 M_\odot$). The precision and accuracy of neutron star mass measurements in spiders are driven by both $i$ and $K_2$. We find a best-fit $i=62.8\pm2.5^\circ$, about $15^\circ$ lower than that reported by Y23, which explains most of the mass discrepancy.
Indeed at lower orbital inclinations, one finds higher $M_1$ for the same $K_2$, as can be seen in Figure \ref{fig:K2iM}. 
%At $64^\circ$, we expect to observe $M_1$ of $2 M_\odot$, as was claimed in \cite{Strader19}.
%
This difference can in principle be due to non-modelled variability (they use observations taken in 2019 February-March in their fits) or to the different model they use \citep[\textsc{PHOEBE},][]{phoebe}.
Their different assumptions can also introduce a systematic uncertainty in $i$: they exclude irradiation, which may lead to a higher $i$ to compensate and keep a similar light curve amplitude. 
While the goodness of fit is not quantified in Y23, we note that their fits perform poorly around the light curve minima.

%Using the widest and most conservative range of values for $i$, taking the uncertainties on $i$ into account, are shown in gray. 

%The range of possible $M_1$ values are between $1.4-3 M_\odot$.

Both our measurement and that of Y23 rely on the $K_2$ reported by \cite{Strader19}.
%
%This is not ideal, since light curves and radial velocity curves should be fit simultaneously.
%
In general $K_2$ measurements can also be affected by systematics when the center of light is displaced from the center of mass of the companion.
To correct for this, \cite{Linares18b} applied an empirical K correction using spectral lines detected from both the day and night sides of the companion of PSR J2215+5135. This allowed a more robust determination of both $K_2$ and $i$ \citep{Linares18b}.
Because we do not have such high-quality spectroscopic data available at present for J1622, we take a conservative uncertainty on $K_2$ by using three times the $1\sigma$ error reported by \cite{Strader19} in our prior. Instead, Y23 used the $1\sigma$ purely statistical uncertainty on $K_2$ as their prior.
% they derive $M_1$ \citep{Y23}. Uncertainties from $K_2$ dominate the uncertainties on $q$, and therefore lead to lower uncertainties on their mass estimates compared to ours.
From our larger uncertainties on $K_2$, which conservatively include systematics, we obtain larger uncertainties on $M_1$ compared to Y23.
Given the low amount of irradiation that we find (Sec.~\ref{sec:irradiation}), however, the K correction should be minor.
%
%and allow us to be confident in our fitted inclination and derived masses. 
High-quality spectroscopy of J1622 along the orbit should reduce the uncertainties in $K_2$ and help determine the mass of the pulsar with better precision.

While we await a more precise dynamical solution, our best-fit value of $2.3\pm0.4\,\text{M}_\odot$ suggests that J1622 \textbf{potentially} hosts a neutron star with one of the highest known masses. The redback PSR J2215+5135 has a similar $M_1$ at $2.27_{-0.16}^{+0.17}\,\text{M}_\odot$ \citep{Linares18b}, while the black widow PSR J0952-0607 has a slightly higher $M_1$ of $2.35 \pm 0.17\,\text{M}_\odot$ \citep[][but the velocities of the night side were ill-constrained in that case]{j0952}. These are two of the most massive millisecond pulsars yet detected. Super-massive neutron stars (with masses above $2$\,M$_{\odot}$) are key to constrain the equation of state of neutron stars, potentially probing the QCD phase diagram. In particular, deconfinement of quarks and phase transitions are impossible to determine solely from radius measurements \citep{qcdrad} and observations of neutron stars with masses above 2 $\text{M}_\odot$ allow many equations of state to be discarded \citep{eosrej}.

\section{Acknowledgments}
We thank the late T. Marsh for the use of {\sc molly}, and acknowledge the use
of data from the UVES Paranal Observatory Project (ESO DDT Program ID
266.D-5655).
We thank J. Strader for kindly providing the SOAR spectra of J1622, J. Casares for discussions on atmospheric lines as temperature
tracers, K. Koljonen for discussion of linked model fits, and T. Shahbaz for advice on MCMC fitting algorithms.
This project has received funding from the European Research Council
(ERC) under the European Union’s Horizon 2020 research and innovation
programme (grant agreements No. 715051 and 101002352). DMS acknowledges support by the Spanish Ministry of Science via the Plan de Generacion de conocimiento PID2020-120323GB-I00 and PID2021-124879NB-I00. VSD and \textsc{ULTRACAM} are supported by the STFC (grant No. ST/Z000033/1).
This work has made use of data from the European Space Agency (ESA) mission
{\it Gaia} (\url{https://www.cosmos.esa.int/gaia}), processed by the {\it Gaia}
Data Processing and Analysis Consortium (DPAC,
\url{https://www.cosmos.esa.int/web/gaia/dpac/consortium}). Funding for the DPAC
has been provided by national institutions, in particular the institutions
participating in the {\it Gaia} Multilateral Agreement.

\section*{Data Availability}
The ULTRACAM light curves presented in this paper are available through a public Zenodo Repository: \url{https://zenodo.org/records/12731300}. The raw images can be requested by contacting Vikram S. Dhillon and Rene P. Breton. The NOT light curves modeled in this paper are available at \url{https://zenodo.org/records/12744062} and the raw images can requested by contacting Marco Turchetta.

\vspace{5mm}
\facilities{NTT:3.5m, NOT:2.56m}

\software{\textsc{ICARUS} \cite{Breton2012},
          \textsc{EMCEE} \citep{FM2019},
          \textsc{DYNESTY} \citep{dynestycode}
          }

\appendix

\section{Corner Plots}\label{corners}
% \begin{figure}[h!]
%     \centering
%     \includegraphics[scale=0.29]{240130_ucam_toptsub_fixi_nooffset.pdf}
%     \caption{This is the corner plot for the ULTRACAM fit using a symmetric model. We consider the effect of irradiation from the pulsar and see that after applying a Gaussian prior on the base temperature and a flat prior on the irradiation temperature, we have a base temperature best-fit value of 6793.17K  with 1899.93K irradiation temperature. In this case, the base temperature is greater than expected from the prior placed upon it.}
%     \label{fig:ucamsymcorner}
% \end{figure}

% \begin{figure}[h!]
%     \centering
%     \includegraphics[scale=0.29]{240131_not_toptsub_fixi_nooffset.pdf}
%     \caption{This is the corner plot for the NOT fit using a symmetric model. We consider the effect of irradiation from the pulsar with the same priors as the fit with the ULTRACAM data. Here, we have a base temperature best-fit value of 6506.93K with 3081.92K irradiation temperature. The base temperature is lower than the one found with the ULTRACAM symmetric model fit, while the irradiation temperature is greater here by over 1000K.}
%     \label{fig:notsymcorner}
% \end{figure}

\begin{figure}[ht!]
    \centering
    \includegraphics[scale=0.23]{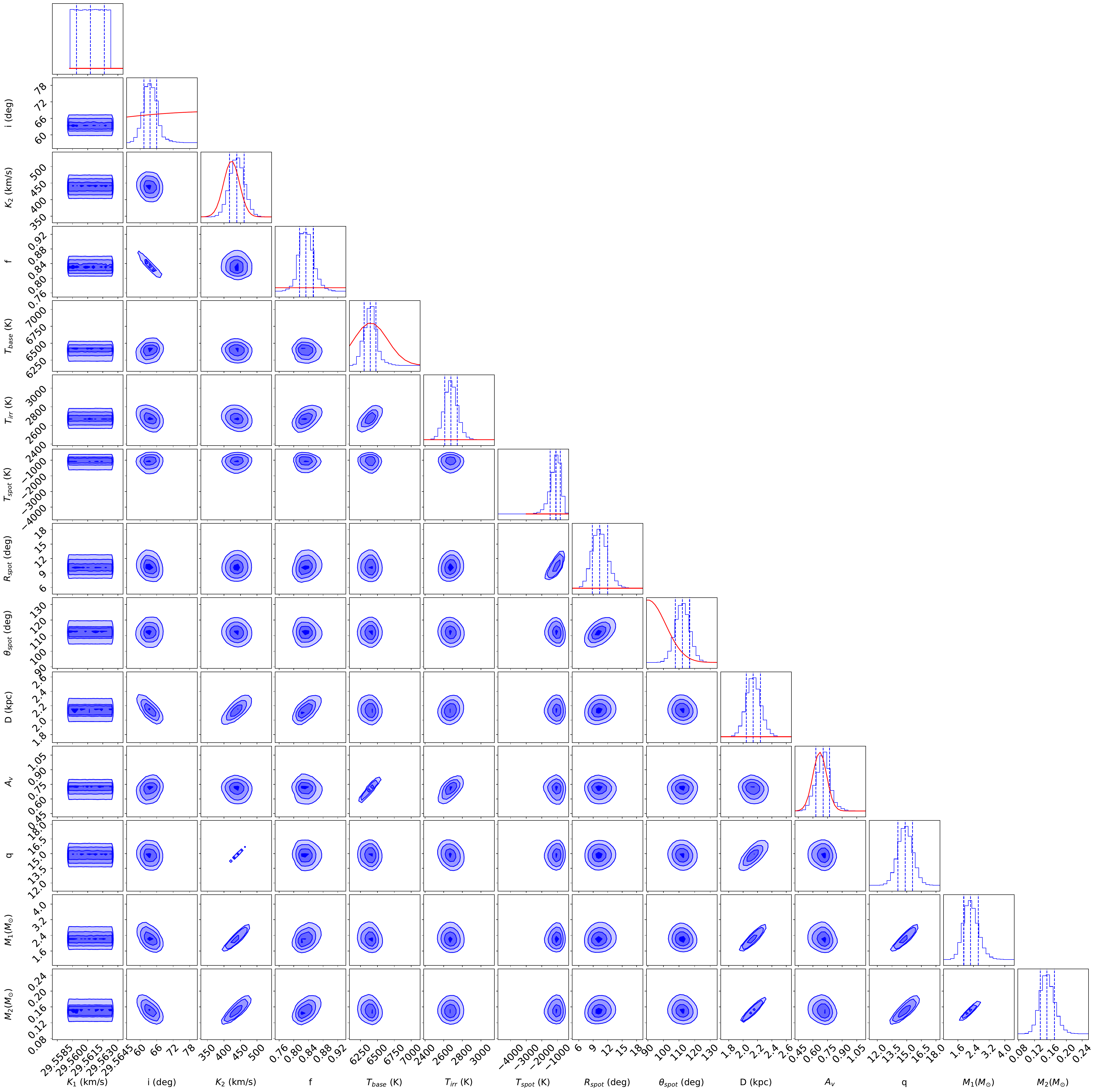}

    \caption{Corner plot for the NTT fit using an asymmetric model. We fit for both irradiation and the temperature and spread of a cold spot placed on the leading side of the star. Contour levels shown are $0.5\sigma$, $1\sigma$, $1.5\sigma$, and $2\sigma$. Prior distributions are in red, while posterior distributions are in blue. See Table \ref{tab:derived} for central values and uncertainties.}
    \label{fig:ucamcscorner}
\end{figure}

\begin{figure}[ht!]
    \centering
    \includegraphics[scale=0.23]{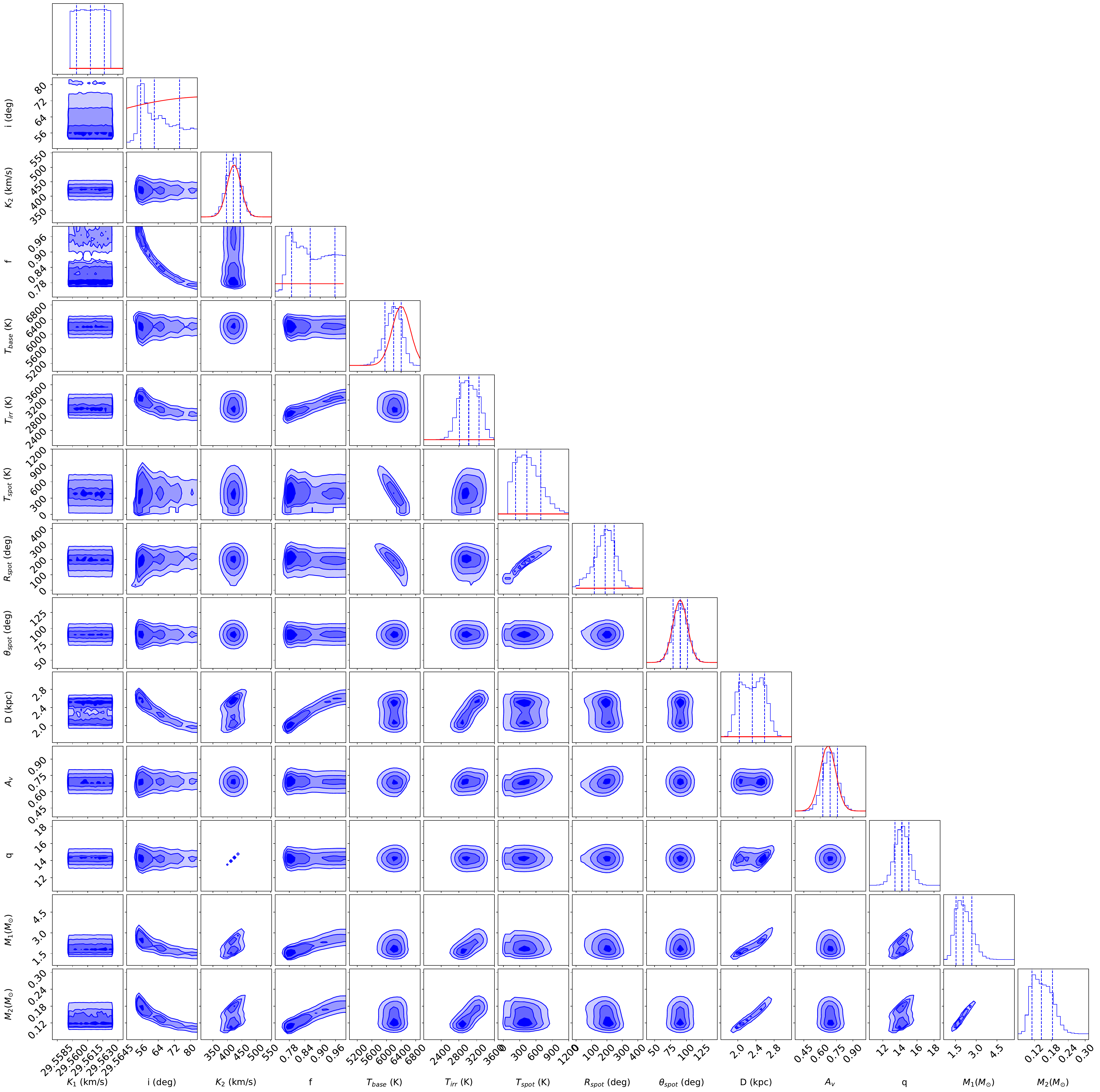}
    
    \caption{Corner plot for the NOT fit using an asymmetric model. We fit for both irradiation and the temperature and spread of a hot spot placed on the leading side of the star. Contour levels shown are $0.5\sigma$, $1\sigma$, $1.5\sigma$, and $2\sigma$. Prior distributions are in red, while posterior distributions are in blue. See Table \ref{tab:derived} for central values and uncertainties.}
    \label{fig:nothscorner}
\end{figure}

\begin{figure}[ht!]
    \centering
    \includegraphics[scale=0.17]{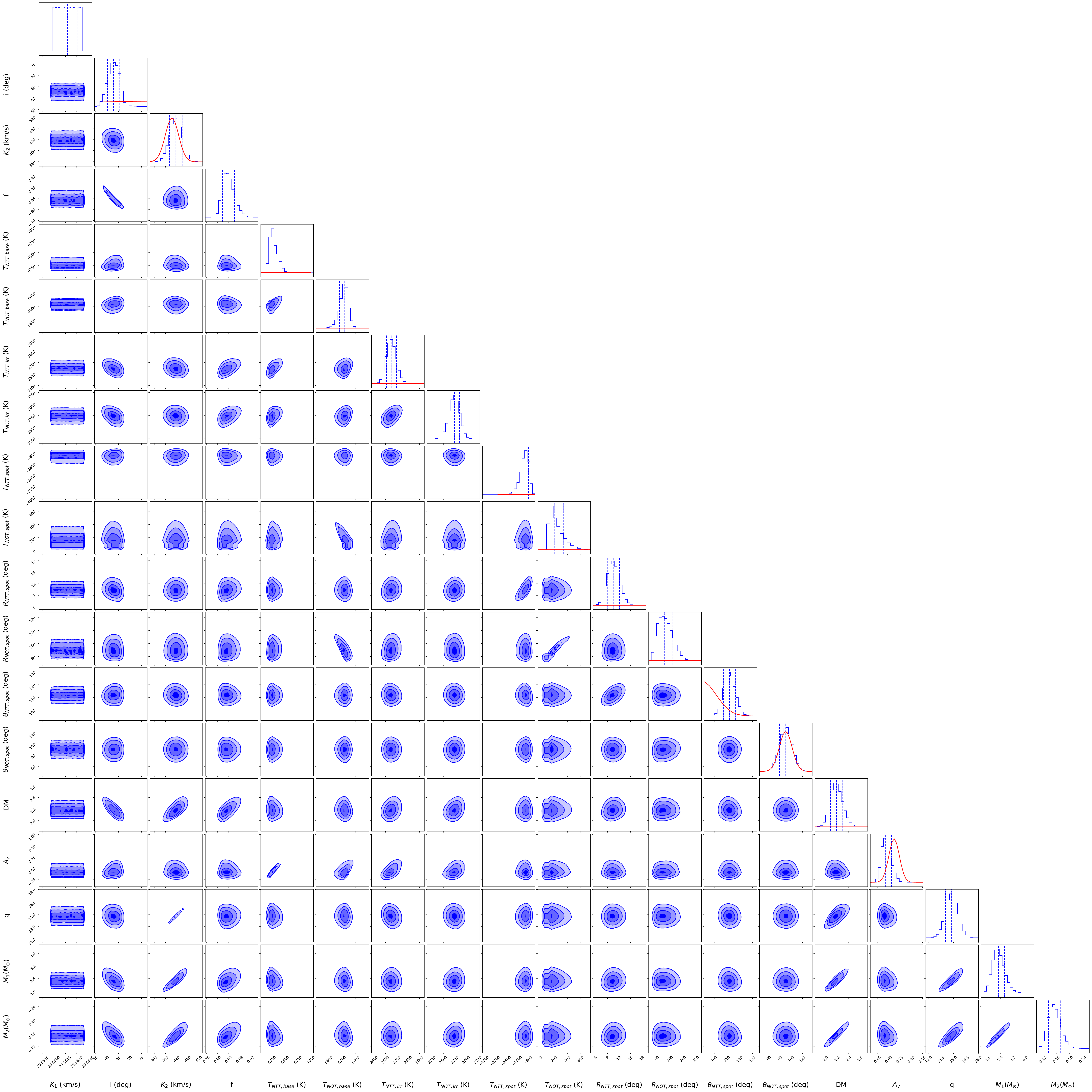}
    \caption{Corner plot for the linked fit with a cold spot for the NTT data and a hot spot for the NOT data. Contour levels shown are $0.5\sigma$, $1\sigma$, $1.5\sigma$, and $2\sigma$. Prior distributions are in red, while posterior distributions are in blue. See Table \ref{tab:derived} for central values and uncertainties.}
    \label{fig:linkedcorner}
\end{figure}

\bibliography{references}{}
\bibliographystyle{aasjournal}

%% This command is needed to show the entire author+affiliation list when
%% the collaboration and author truncation commands are used.  It has to
%% go at the end of the manuscript.
%\allauthors

%% Include this line if you are using the \added, \replaced, \deleted
%% commands to see a summary list of all changes at the end of the article.
%\listofchanges

\end{document}